\documentclass[a4paper,twoside,twocolumn,english,showpacs,superscriptaddress]{revtex4}
\usepackage{lmodern}

\usepackage[T1]{fontenc}
\usepackage[utf8]{luainputenc}
\usepackage{babel}
\usepackage{amsmath}
\usepackage{amssymb}
\usepackage{graphicx}
\usepackage{esint}
\usepackage[unicode=true,pdfusetitle,
 bookmarks=true,bookmarksnumbered=false,bookmarksopen=false,
 breaklinks=false,pdfborder={0 0 1},backref=false,colorlinks=false]
 {hyperref}

\makeatletter

\pdfpageheight\paperheight
\pdfpagewidth\paperwidth

\@ifundefined{textcolor}{}
{%
 \definecolor{BLACK}{gray}{0}
 \definecolor{WHITE}{gray}{1}
 \definecolor{RED}{rgb}{1,0,0}
 \definecolor{GREEN}{rgb}{0,1,0}
 \definecolor{BLUE}{rgb}{0,0,1}
 \definecolor{CYAN}{cmyk}{1,0,0,0}
 \definecolor{MAGENTA}{cmyk}{0,1,0,0}
 \definecolor{YELLOW}{cmyk}{0,0,1,0}
}

\usepackage{babel}
\usepackage{babel}

\usepackage{graphics}
\usepackage{epsfig}
\usepackage{epsf}

\makeatother

\begin{document}

\title{Dirty bosons in a three-dimensional harmonic trap }

\author{Tama Khellil}

\email{khellil.lpth@gmail.com}

\affiliation{Insitut f\"ur Theoretische Physik, Freie Universit\"{a}t Berlin, Arnimallee
14, 14195 Berlin, Germany}

\author{Axel Pelster}

\email{axel.pelster@physik.uni-kl.de}

\affiliation{Fachbereich Physik und Forschungszentrum OPTIMAS, Technische Universit\"{a}t
Kaiserslautern, 67663 Kaiserslautern, Germany}
\begin{abstract}
We study a three-dimensional Bose-Einstein condensate in an isotropic
harmonic trapping potential with an additional delta-correlated disorder
potential at both zero and finite temperature and investigate the
emergence of a Bose-glass phase for increasing disorder strength.
To this end, we revisit a quite recent non-perturbative approach towards
the dirty boson problem, which relies on the Hartree-Fock mean-field
theory and is worked out on the basis of the replica method, and extend
it from the homogeneous case to a harmonic confinement. At first,
we solve the zero-temperature self-consistency equations for the respective
density contributions, which are obtained via the Hartree-Fock theory
within the Thomas-Fermi approximation. Additionally we use a variational
ansatz, whose results turn out to coincide qualitatively with those
obtained from the Thomas-Fermi approximation. In particular, a first-order
quantum phase transition from the superfluid phase to the Bose-glass
phase is detected at a critical disorder strength, which agrees with
findings in the literature.  Afterwards, we consider the three-dimensional
dirty boson problem at finite temperature. This allows us to study
the impact of both temperature and disorder fluctuations on the respective
components of the density as well as their Thomas-Fermi radii. In
particular, we find that a superfluid region, a Bose-glass region,
and a thermal region coexist for smaller disorder strengths. Furthermore,
depending on the respective system parameters, three phase transitions
are detected, namely, one from the superfluid to the Bose-glass phase,
another one from the Bose-glass to the thermal phase, and finally
one from the superfluid to the thermal phase. 
\end{abstract}


\maketitle

\section{Introduction}

The combined effect of disorder and two-particle interactions in the
dirty boson problem yields a competition between localization and
superfluidity \cite{Intro-11}. Experimentally, the dirty boson problem
was first studied with superfluid helium in porous media like aerosol
glasses (Vycor), where the pores are modeled by statistically distributed
local scatterers \cite{Intro-91,Intro-92,Intro-93,Intro-94}. Disorder
in Bose gases appears either naturally as, e.g., in magnetic wire
traps \cite{Intro-75,Intro-76,Intro-77,Fortagh,Schmiedmayer}, where
imperfections of the wire itself can induce local disorder, or it
may be created artificially and controllably as, e.g., by using laser
speckle fields \cite{Intro-78,Intro-79,Intro-21,Goodmann,Intro-16}.
A set-up more in the spirit of condensed matter physics relies on
a Bose gas with impurity atoms of another species trapped in a deep
optical lattice, so the latter represent randomly distributed scatterers
\cite{Intro-13,Intro-14}. Furthermore, an incommensurate optical
lattice can provide a pseudo-random potential for an ultracold Bose
gas \cite{Lewenstein,Ertmer,intro-17}. 

The homogeneous dirty boson model is important as it provides a good
description at the center of a harmonic trap and, thus, serves as
a starting point for treating a harmonic confinement within the Thomas-Fermi
approximation. Furthermore, recently it has even become possible to
experimentally realize box-like traps \cite{Hadzibabic}, which approximate
the homogeneous case in the thermodynamic limit. The first important
theoretical result for the homogeneous dirty boson was obtained by
Huang and Meng, who found, within the Bogoliubov theory \cite{Intro-7},
that a weak disorder potential with delta correlation leads to a depletion
of both the condensate and the superfluid density due to the localization
of bosons in the respective minima of the random potential \cite{HM-1}.
Later on their theory was extended in different research directions.
Results for the shift of the velocity of sound as well as for its
damping due to collisions with the external field are worked out in
Ref.~\cite{HM-3}. Furthermore, the delta-correlated random potential
was generalized to experimentally more realistic disorder correlations
with a finite correlation length, e.g., a Gaussian correlation was
discussed in Ref.~\cite{HM-4} and laser speckles are treated at
zero \cite{HM-6} and finite temperature \cite{HM-7}. Also the disorder-induced
shift of the critical temperature was analyzed in Refs.~\cite{HM-5,HM-2}.
Furthermore, it was shown that dirty dipolar Bose gases yield characteristic
directional dependences for thermodynamic quantities due to the emerging
anisotropy of superfluidity at zero \cite{HM-8,HM-9} and finite temperature
\cite{HM-10,Boujam3a1,Boujam3a2}. The location of superfluid, Bose-glass,
and normal phase in the phase diagram spanned by disorder strength
and temperature was qualitatively analyzed for the first time in Ref.~\cite{Intro-90}
on the basis of a Hartree-Fock mean-field theory with the replica
method. In addition, increasing the disorder strength at small temperatures
yields a first-order quantum phase transition from a superfluid to
a Bose-glass phase, where in the latter case all particles reside
in the respective minima of the random potential. This prediction
is achieved at zero temperature by solving the underlying Gross-Pitaevskii
equation with a random phase approximation \cite{Navez}, as well
as at finite temperature by a stochastic self-consistent mean-field
approach using two chemical potentials, one for the condensate and
one for the excited particles \cite{Yukalov1}. Numerically, Monte-Carlo
(MC) simulations have been applied to study the homogeneous dirty
boson problem. For instance, diffusion MC in Ref.~\cite{Astrakharchik}
obtained the surprising result that at zero temperature a strong enough
disorder yields a superfluid density, which is larger than the condensate
density. Furthermore, worm-algorithm MC was able to determine the
dynamic critical exponent of the quantum phase transition from the
Bose-glass to the superfluid in two dimensions at zero \cite{Monte-Carlo1}
and finite temperature \cite{Monte-Carlo2}.

Adding a harmonic trap to the dirty Bose gas problem makes it realistic
but more complicated to treat than the homogeneous one. Since the
collective excitation frequencies of harmonically trapped bosons can
be measured very accurately, their change due to disorder was investigated
in Ref.~\cite{Collective-excitations} at zero temperature. As the
collective excitation frequencies turn out to decrease rapidly with
the correlation length of disorder, one would have to reduce the correlation
length of the laser speckles in Ref.~\cite{Intro-79} from $10{\rm \,{\mu m}}$
by a factor of $10$ in order to be able to detect any shift due to
the disorder. The expansion of a Bose-Einstein condensate (BEC) at
zero temperature in the presence of a random potential was studied
in Ref.~\cite{Boris-trap}. Depending on the strength of disorder
and the two-particle interaction, a crossover from localization to
diffusion was observed. The shape and size of the local minicondensates
in the disorder landscape were investigated energetically at zero
temperature in Refs.~\cite{Nattermann,Natterman2}, where it was
deduced that, for decreasing disorder strength, the Bose-glass phase
becomes unstable and goes over into the superfluid. At finite temperature
the disorder-induced shift of the critical temperature was analyzed
for a harmonic confinement in Ref.~\cite{Timmer}. The impact of
the random potential upon the quantum fluctuations at finite temperature
was also studied in Refs.~\cite{Gaul-Muller-1,Gaul-Muller-2}. Furthermore,
based on Ref.~\cite{Intro-90}, Ref.~\cite{Paper2} worked out in
detail a non-perturbative approach to the dirty boson problem, which
relies on the Hartree-Fock theory and the Parisi replica method, for
a weakly interacting Bose-gas within a harmonic confinement and a
delta-correlated disorder potential at finite temperature. Its application
to a quasi one-dimensional BEC at zero temperature \cite{Paper1}
reveals a redistribution of the minicondensates from the edge of the
atomic cloud to the trap center for increasing disorder strengths.
Despite all these many theoretical predictions, so far no experiment
has tested them quantitatively.

In the present paper we treat analytically the problem of a three-dimensional
trapped BEC in a disorder potential on the basis of Ref.~\cite{Paper2}.
To this end, we start by describing the underlying dirty boson model
and developing a Hartree-Fock mean-field theory in Section II. Then
we treat, as a first step, the zero-temperature case in Section III,
which allows us to study the impact of the disorder on the distribution
of the condensate density and the Bose-glass order parameter, which
quantifies the density of the bosons in the local minima of the disorder
potential. We deal first with the simpler homogeneous case, and then
we analyze the isotropic harmonically trapped one. Using the corresponding
self-consistency equations obtained via the Hartree-Fock mean-field
theory, we investigate within the Thomas-Fermi approximation the existence
of the Bose-glass phase. We additionally use a variational ansatz,
whose results turn out to coincide qualitatively with the ones obtained
via the Thomas-Fermi approximation. In Section IV we consider the
three-dimensional dirty BEC system to be at finite temperature. We
restrict ourselves first to the homogeneous dirty case, after that
to the trapped clean case. Afterwards we treat the trapped disordered
case at finite temperature using the Thomas-Fermi approximation. This
allows us to study the impact of both temperature and disorder fluctuations
on the respective components of the density as well as their Thomas-Fermi
radii. In particular, three regions coexist, namely, a superfluid
region, a Bose-glass region, and a thermal region. Furthermore, depending
on the respective system parameters, three phase transitions are detected,
one from the superfluid to the Bose-glass phase, another one from
the Bose-glass to the thermal phase, where all bosons are in the excited
states, and a third one from the superfluid to the thermal phase.

\section{Hartree-Fock Mean-Field Theory in 3D }

The model of a three-dimensional weakly interacting homogeneous Bose
gas in a delta-correlated disorder potential was studied within the
Hartree-Fock mean-field theory in Ref.~\cite{Intro-90} by applying
the Parisi replica method \cite{Intro-36,Intro-37,Intro-33}. This
Hartree-Fock theory is extended in Ref.~\cite{Paper2} to a harmonic
confinement. Let us briefly summarize the main result of Ref.~\cite{Paper2},
which relies on deriving a semiclassical approximation for the underlying
free energy.

We consider a three-dimensional Bose gas in an isotropic harmonic
potential $V(\mathbf{r})=M\Omega^{2}\mathbf{r}^{2}/2$ with the trap
frequency $\Omega$, the particle mass $M$ and the contact interaction
potential $V^{({\rm int})}(\mathbf{r}-\mathbf{r}')=g\delta(\mathbf{r}-\mathbf{r}')$.
The interaction coupling strength $g=4\pi\hslash^{2}a/M$ depends
on the s-wave scattering length $a$, which has to be positive in
order to obtain a stable BEC. We assume for the disorder potential
$U(\mathbf{r})$ that it is homogeneous after performing the disorder
ensemble average, denoted by $\overline{\bullet}$, over all possible
realizations. Thus, the expectation value of the disorder potential
can be set to vanish without loss of generality,

\begin{eqnarray}
\overline{U(\mathbf{r})}=0,\label{ZP1}
\end{eqnarray}
and its correlation function is assumed to be proportional to a delta-function,
\begin{eqnarray}
\overline{U(\mathbf{r}_{1})U(\mathbf{r}_{2})}=D\,\delta(\mathbf{r}_{1}-\mathbf{r}_{2})\,,\label{ZP2}
\end{eqnarray}
where $D$ denotes the disorder strength.

By working out the Hartree-Fock mean-field theory within the replica
method, Ref.~\cite{Paper2} obtains self-consistency equations, which
determine the particle density $n(\mathbf{r})$ as well as the order
parameter of the superfluid $n_{0}(\mathbf{r})$, representing the
condensate density, the order parameter of the Bose-glass phase $q(\mathbf{r})$
defined in Ref.~\cite{Intro-90}, that stands for the density of
the particles being condensed in the respective minima of the disorder
potential, and $n_{{\rm th}}\left(\mathbf{r}\right)$, which represents
the density of the particles in the excited states. The Hartree-Fock
mean-field theory with the help of the replica method and a semiclassical
approximation leads to the free energy \cite{Paper2}: 
\begin{alignat}{1}
\mathcal{F} & =4\pi\!\int_{0}^{\infty}drr^{2}\Biggl\{-g\left[q(r)+n_{0}(r)+n_{{\rm th}}\left(r\right)\right]^{2}-\frac{g}{2}n_{0}^{2}(r)\nonumber \\
 & \!\!\!\!\!\!+\frac{D}{\hbar}Q_{0}(r)\left[q(r)+n_{0}(r)+n_{{\rm th}}\left(r\right)\right]-\sqrt{n_{0}(r)}\nonumber \\
 & \!\!\!\!\!\!\times\Biggl\{\mu+\frac{\hbar^{2}}{2M}\frac{1}{r^{2}}\frac{\partial}{\partial r}\left(r^{2}\frac{\partial}{\partial r}\right)-2g\left[q(r)+n_{0}(r)+n_{{\rm th}}\left(r\right)\right]\nonumber \\
 & \!\!\!\!\!\!-V(r)+\frac{D}{\hbar}Q_{0}(r)\Biggr\}\sqrt{n_{0}(r)}-2D\sqrt{\pi}\left(\frac{M}{2\pi\hbar^{2}}\right)^{3/2}\label{F1}\\
 & \!\!\!\!\!\!\times\left[q(r)+n_{0}(r)+n_{{\rm th}}\left(r\right)\right]-\frac{1}{\beta}\left(\frac{M}{2\pi\hbar^{2}\beta}\right)^{3/2}\zeta_{5/2}\left(e^{\beta\mu_{r}(r)}\right)\nonumber \\
 & \!\!\!\!\!\!\left.\times\sqrt{-\mu+2g\left[q(r)+n_{0}(r)+n_{{\rm th}}\left(r\right)\right]+V(r)-\frac{D}{\hbar}Q_{0}(r)}\right\} .\nonumber 
\end{alignat}
Here all functions only depend on the radial coordinate $r=\left|\mathbf{r}\right|$
due to the assumed spatial isotropy. Furthermore, $\mu$ denotes the
chemical potential, $\mu_{r}(r)=\mu-V(r)-2g\left[q(r)+n_{0}(r)+n_{{\rm th}}\left(r\right)\right]-\pi D^{2}\left(\frac{M}{2\pi\hbar^{2}}\right)^{3}$
represents the renormalized chemical potential, and $Q_{0}(r)$ stands
for an auxiliary function, which appears within the Hartree-Fock theory:
\begin{alignat}{1}
Q_{0}(r) & =-2\sqrt{\pi}\hbar{\displaystyle \left(\frac{M}{2\pi\hbar^{2}}\right)^{3/2}}\nonumber \\
 & \times\left[\sqrt{\pi}D{\displaystyle \left(\frac{M}{2\pi\hbar^{2}}\right)^{3/2}}+\sqrt{-\mu_{r}(r)}\right].\label{Q3}
\end{alignat}
From the thermodynamic relation $N=-\frac{\partial\mathcal{F}}{\partial\mu}$
we obtain 
\begin{equation}
N=4\pi\int_{0}^{\infty}r^{2}n(r)dr,\label{7-1}
\end{equation}
which defines the particle density $n(r)$.

Extremising the free energy \eqref{F1} with respect to the functions
$n_{0}(r)$, $q(r)$, $n_{{\rm th}}\left(r\right),$ and $Q_{0}(r)$,
i.e., $\frac{\delta\mathcal{F}}{\delta n{}_{0}(r')}=0$, $\frac{\delta\mathcal{F}}{\delta q(r')}=0$,
$\frac{\delta\mathcal{F}}{\delta n_{{\rm th}}(r')}=0$, and $\frac{\delta\mathcal{F}}{\delta Q_{0}(r')}=0$,
respectively, yields, together with Eq. \eqref{7-1}, four coupled
self-consistency equations between the respective density contributions:
a nonlinear differential equation for the condensate density $n_{0}(r)$,
\begin{alignat}{1}
\Biggl\{-gn{}_{0}(r)+\left[\sqrt{-\mu+d^{2}+2gn(r)+V(r)}+d\right]^{2}\label{6-2}\\
\!\!\!\!\!\!\!\!\!\!\!\!\!\!\!-\frac{\hslash^{2}}{2M}\frac{1}{r^{2}}\frac{\partial}{\partial r}\left(r^{2}\frac{\partial}{\partial r}\right)\Biggr\}\sqrt{n{}_{0}(r)} & =0,\nonumber 
\end{alignat}
an algebraic equation for the Bose-glass order parameter $q(r)$,

\begin{equation}
q(r)=\frac{dn{}_{0}(r)}{\sqrt{-\mu+d^{2}+2gn(r)+V(r)}},\label{5-1}
\end{equation}
the thermal density $n_{{\rm th}}\left(r\right)$,

\begin{equation}
n_{{\rm th}}\left(r\right)=\left(\frac{M}{2\pi\beta\hslash^{2}}\right)^{3/2}\varsigma_{\ 3/2}\left(e^{\beta\ [\mu-d^{2}-2gn(r)-V(r)]}\right),\label{3-1}
\end{equation}
with the polylogarithmic function $\zeta_{\nu}(z)=\sum_{\mathtt{n}=1}^{\infty}\frac{z^{\mathtt{n}}}{\mathtt{n}^{\nu}}$,
and the sum of the above three densities, which turns out to be the
total density $n(r)$, 
\begin{equation}
n(r)=n_{0}(r)+q(r)+n_{{\rm th}}\left(r\right),\label{n-3}
\end{equation}
where $d=\sqrt{\pi}D\left(M/2\pi\hslash^{2}\right)^{3/2}$characterizes
the disorder strength. 

In the following we deal first with the zero-temperature Bose gas,
then we treat the finite-temperature case via the Thomas-Fermi approximation.

\section{3D Dirty Bosons at Zero Temperature }

In this section we consider the three-dimensional dirty BEC system
at zero temperature, where the thermal density vanishes, i.e., $n_{{\rm th}}\left(r\right)=0$.

\subsection{Homogeneous case}

\begin{figure}[t]
\includegraphics[width=0.45\textwidth]{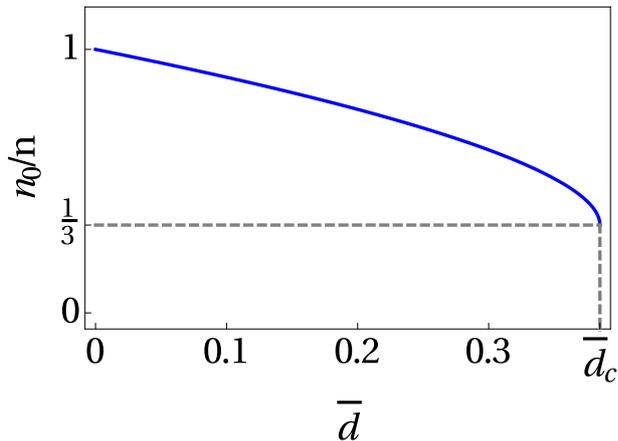} \protect\caption{\label{fig:Homo-3d}(Color online) Condensate fraction $n_{0}/n$
as function of dimensionless disorder strength $\overline{d}$. }
\end{figure}

We start with the homogeneous case since it is the simplest one, where
in the absence of the trap we have $V(r)=0$. At zero temperature
we only need Eqs. \eqref{6-2}, \eqref{5-1}, and \eqref{n-3}, which
reduce in the superfluid phase to:

\begin{equation}
gn{}_{0}=\left(\sqrt{-\mu+d^{2}+2gn}+d\right)^{2},\label{6-3}
\end{equation}

\begin{equation}
q=\frac{dn{}_{0}}{\sqrt{-\mu+d^{2}+2gn}},\label{5-2}
\end{equation}
\begin{equation}
n=n_{0}+q.\label{n-1}
\end{equation}
Note that we dropped here the spatial dependency of all densities
due to the homogeneity. From Eqs. \eqref{6-3}--\eqref{n-1} we get
the following algebraic third-order equation for determining the condensate
fraction $n{}_{0}/n$:

\begin{equation}
\left(\frac{n_{0}}{n}\right)^{3/2}-\sqrt{\frac{n_{0}}{n}}+\overline{d}=0.\label{Homo-3D}
\end{equation}
Here $\overline{d}=\frac{\xi}{\mathcal{L}}$ denotes the dimensionless
disorder strength, where $\xi=\frac{\hbar}{\sqrt{2Mgn}}$ stands for
the coherence length, and $\mathcal{L}=\frac{2\pi\hbar^{4}}{M^{2}D}$
represents the Larkin length, which characterizes the strength of
disorder \cite{Larkin,Nattermann}. Figure~\ref{fig:Homo-3d} predicts
that the equation for condensate density does not have a solution
after the critical value $\overline{d}_{{\rm {c}}}=\sqrt{\frac{1}{3}}-\left(\frac{1}{3}\right)^{3/2}\simeq0.384$.
We interpret this as a sign that a first-order quantum phase transition
occurs in the homogeneous case from the superfluid phase, where the
particles are either condensed or in the local minima of the disorder,
to the Bose-glass phase, where there is no condensate at all and all
bosons are localized in the minima of the disorder potential. This
suggests that a quantum phase transition will also appear in the trapped
case, which is studied later on in Subsection III.C. 

Now we check whether our results are compatible with the Huang-Meng
theory \cite{HM-1,HM-3,HM-4,HM-5,HM-2}, where the Bose-glass order
parameter  of a homogeneous dilute Bose gas at zero temperature in
case of weak disorder regime is deduced within the seminal Bogoliubov
theory. The Bose-glass order parameter in three dimensions via the
Huang-Meng theory is proportional to the disorder strength and yields
in dimensionless form: 
\begin{equation}
\frac{q_{{\rm HM}}}{\sqrt{n/g}}=\frac{\overline{d}}{\sqrt{2}}.\label{3.23-1}
\end{equation}

In our Hartree-Fock mean-field theory the Bose-glass order parameter
in case of weak disorder strength turns out to be

\begin{equation}
\ensuremath{\frac{q_{w}}{\sqrt{n/g}}=\overline{d}.}\label{19-1-1}
\end{equation}

Thus, our theory agrees with the Huang-Meng theory at least qualitatively.
But quantitatively the comparison of Eqs. \eqref{3.23-1} and \eqref{19-1-1}
reveals that a factor of $\sqrt{2}$ is missing in our result \eqref{19-1-1}.
This is due to the fact that the Hartree-Fock theory does not contain
the Bogoliubov channel, which is included in the Huang-Meng theory.

According to Ref.~\cite{Navez}, the disorder strength value corresponding
to the quantum phase transition is $\overline{d}_{{\rm {c}}}=0.53$.
Thus, our quantum phase transition disorder value $\overline{d}_{{\rm {c}}}=0.384$
is of the same order as the one in Ref.~\cite{Navez}, but again
we miss a factor of $\sqrt{2}$ in our result. In the one-dimensional
case, as discussed in Ref.~\cite{Paper1}, a factor $2^{3/2}$ is
missing, while in the three-dimensional case the discrepancy only
amounts to a factor of $\sqrt{2}$, so we conclude that our Hartree-Fock
theory is more compatible with the literature in higher dimensions
than in lower ones.

Furthermore, we compare the critical value of the disorder strength
$\overline{d}_{{\rm {c}}}$ with the non-perturbative approach of
Refs.~\cite{Nattermann,Natterman2}, which starts from the Bose-glass
phase and goes towards the superfluid phase for decreasing disorder
strength. By investigating energetically shape and size of the local
minicondensates in the disorder landscape, the quantum phase transition
is predicted to occur at the disorder strength value $\tilde{d}=\sqrt{\frac{3}{8\pi}}\simeq0.345$,
which is again of the same order as our $\tilde{d}_{{\rm c}}$.

\subsection{Thomas-Fermi approximation}

We deal now with the trapped case. The exact analytical solution of
the differential equation \eqref{6-2} is impossible to obtain even
in the absence of disorder. Therefore, we approximate its solution
via the Thomas-Fermi (TF) approximation, which is based on neglecting
the kinetic energy.

It turns out that we have to distinguish between two different spatial
regions: the superfluid region, where the bosons are distributed in
the condensate as well as in the minima of the disorder potential,
and the Bose-glass region, where there are no bosons in the global
condensate and all bosons contribute only to the local Bose-Einstein
condensates. In the following the radius of the superfluid region,
i.e., the condensate radius, is denoted by $R_{{\rm {TF1}}}$, while
the radius of the whole bosonic cloud $R_{{\rm {TF2}}}$ is called
the cloud radius.

Within the TF approximation the algebraic equations \eqref{5-1} and
\eqref{n-3} remain the same, but the differential equation \eqref{6-2}
reduces to an algebraic relation in the superfluid region: 
\begin{equation}
gn{}_{0}(r)=\left[\sqrt{-\mu+d^{2}+2gn(r)+V(r)}+d\right]^{2}.\label{6-4-1}
\end{equation}
Outside the superfluid region, i.e., in the Bose-glass region, Eq.
\eqref{6-2} reduces simply to $n{}_{0}(r)=0$. The advantage of the
TF approximation is that now we have only three coupled algebraic
equations.

At first we consider the superfluid region. Equations \eqref{5-1},
\eqref{n-3}, and \eqref{6-4-1} reduce in the superfluid region to: 

\begin{equation}
\tilde{n}_{0}\left(\tilde{r}\right)=\left[\sqrt{-\tilde{\mu}+2\tilde{n}(\tilde{r})+\tilde{r}^{2}}+\tilde{d}\right]^{2},\label{15}
\end{equation}

\begin{equation}
\tilde{q}\left(\tilde{r}\right)=\frac{\tilde{d}\left[\sqrt{-\tilde{\mu}+2\tilde{n}(\tilde{r})+\tilde{r}^{2}}+\tilde{d}\right]^{2}}{\sqrt{-\tilde{\mu}+2\tilde{n}(\tilde{r})+\tilde{r}^{2}}},\label{16}
\end{equation}

\begin{align}
\tilde{n}\left(\tilde{r}\right)=\frac{\left[\sqrt{-\tilde{\mu}+2\tilde{n}(\tilde{r})+\tilde{r}^{2}}+\tilde{d}\right]^{3}}{\sqrt{-\tilde{\mu}+2\tilde{n}(\tilde{r})+\tilde{r}^{2}}},\label{17}
\end{align}
where $\tilde{n}_{0}(\tilde{r})=n_{0}(r)/\overline{n},$ $\tilde{q}(\tilde{r})=q(r)/\overline{n}$,
and $\tilde{n}(\tilde{r})=n(r)/\overline{n}$ denote the dimensionless
condensate density, Bose-glass order parameter, and total density,
respectively. Furthermore, we have introduced the dimensionless radial
coordinate $\tilde{r}=r/R_{{\rm {TF}}}$, the dimensionless chemical
potential $\tilde{\mu}=(\mu-d^{2})/\bar{\mu}$, the dimensionless
disorder strength $\tilde{d}=\frac{\xi}{\mathcal{L}}$, the coherence
length in the center of the trap $\xi=\frac{l^{2}}{R_{{\rm {TF}}}}$,
the oscillator length $l=\sqrt{\frac{\hbar}{M\Omega}}$, the maximal
total density in the clean case $\overline{n}=\bar{\mu}/g$, and the
TF cloud radius $R_{{\rm {TF}}}=l\sqrt{2\bar{\mu}/\hbar\Omega}$.
The chemical potential in the absence of the disorder $\bar{\mu}=\frac{15^{2/5}}{2}\Big(\frac{aN}{l}\Big)^{2/5}\hslash\Omega$
serves here as the underlying energy scale and is deduced from the
normalization condition \eqref{7-1} in the clean case, i.e., for
$d=0$.

Equation \eqref{17} is of the third order with respect to the expression
$\sqrt{-\tilde{\mu}+2\tilde{n}(\tilde{r})+\tilde{r}^{2}}$, therefore,
we use the Cardan method to solve it analytically \cite{Greiner}.
We determine the condensate radius $\tilde{R}_{{\rm TF1}}$ at the
coordinate where the solution of \eqref{17} for the total density
stops to exist, then select the smallest solution, which corresponds
to $\tilde{R}_{{\rm TF1}}=\sqrt{\tilde{\mu}-3\tilde{d}^{2}-6\sqrt{3}\tilde{d}^{2}\cos\left(\frac{\pi}{18}\right)}$.
Now we have just to insert the obtained particle density $\tilde{n}(\tilde{r})$
into the two other equations $(\ref{15})$ and $(\ref{16})$ in order
to get both the condensate density $\tilde{n}_{0}(\tilde{r})$ and
the Bose-glass order parameter  $\tilde{q}(\tilde{r})$, respectively.

In the Bose-glass region the condensate vanishes, i.e., $\tilde{n}_{0}(\tilde{r})=0$
and $\tilde{n}(\tilde{r})=\tilde{q}(\tilde{r})$, and the self-consistency
equation (\ref{5-1}) reduces to:
\begin{equation}
\tilde{q}(\tilde{r})=\frac{\tilde{\mu}-\tilde{r}^{2}}{2}.\label{14}
\end{equation}
This Bose-glass region ends at the cloud radius $\tilde{R}_{{\rm TF2}}=\sqrt{\tilde{\mu}}$.
We also need to write down the dimensionless equivalent of the normalization
condition \eqref{7-1}, which reads: 
\begin{equation}
\int_{0}^{\tilde{R}_{{\rm {TF2}}}}\tilde{n}(\tilde{r})\tilde{r}^{2}d\tilde{r}=\frac{2}{15},\label{18}
\end{equation}
where the total density $\tilde{n}(\tilde{r})$ in equation \eqref{18}
is the combination of the total densities from both the superfluid
region and the Bose-glass region. The purpose of Eq.~\eqref{18}
is to determine the dimensionless chemical potential $\tilde{\mu}$
from the respective system parameters.

\subsection{Thomas-Fermi results}

Before choosing any parameters for the BEC system, we have to justify
using the TF approximation and determine its range of validity. To
this end we rewrite Eq.~\eqref{6-2} in the clean case, where the
total density coincides with the condensate one:

\begin{alignat}{1}
\left[-1+\tilde{n}(\tilde{r})+\tilde{r}^{2}-\left(\frac{\xi}{R_{{\rm {TF}}}}\right)^{2}\frac{1}{\tilde{r}^{2}}\frac{\partial}{\partial\tilde{r}}\left(\tilde{r}^{2}\frac{\partial}{\partial\tilde{r}}\right)\right]\,\nonumber \\
\!\!\!\!\!\!\!\!\!\!\!\!\!\!\!\!\!\!\!\!\!\!\!\!\!\!\!\!\!\!\!\!\!\!\!\!\times\sqrt{\tilde{n}\left(\tilde{r}\right)}=0.\label{PMM-1-1-3-1}
\end{alignat}
We read off from Eq.~\eqref{PMM-1-1-3-1} that the TF approximation
is only justified when $\xi\ll R_{{\rm {TF}}}$.

In this subsection we perform our study for $^{87}{\rm {Rb}}$ atoms
and for the following experimentally realistic parameters: $N=10^{6}$,
$\Omega=2\pi\times100{\rm \,{Hz}}$, and $a=5.29\,{\rm {nm}}$. For
those parameters the oscillator length reads $l=1.08{\rm \,{\mu m}},$
the coherence length turns out to be $\xi=115\,{\rm {nm}},$ and the
Thomas-Fermi radius is given by $R_{{\rm {TF}}}=10.21{\rm \,{\mu m}}$.
Thus the assumption $\xi\ll R_{{\rm {TF}}}$ for the TF approximation
is, indeed, fulfilled.

\begin{widetext}

\begin{figure}[t]
\includegraphics[width=0.8\textwidth]{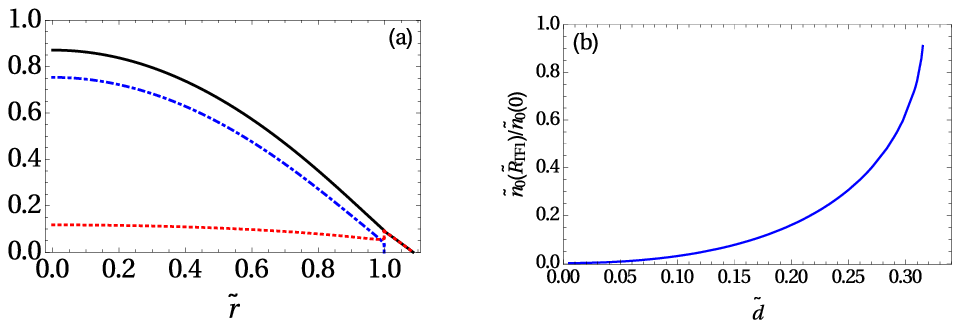}\protect\caption{\label{TF-Iso}(Color online) (a) Total density $\tilde{n}(\tilde{r})$
(solid, black), condensate density $\ensuremath{\tilde{n}_{0}(\tilde{r})}$
(dotted, blue), Bose-glass order parameter $\ensuremath{\tilde{q}(\tilde{r})}$
(dotted-dashed, red) as a function of radial coordinate $\ensuremath{\tilde{r}}$
for the disorder strength $\ensuremath{\tilde{d}=0.175}$ both for
superfluid region and Bose-glass region. (b) Ratio of $\tilde{n}_{0}(\tilde{R}_{{\rm TF1}})$
and $\tilde{n}_{0}(0)$ as a function of disorder strength $\ensuremath{\tilde{d}}$.}
\bigskip{}
\includegraphics[width=0.8\textwidth]{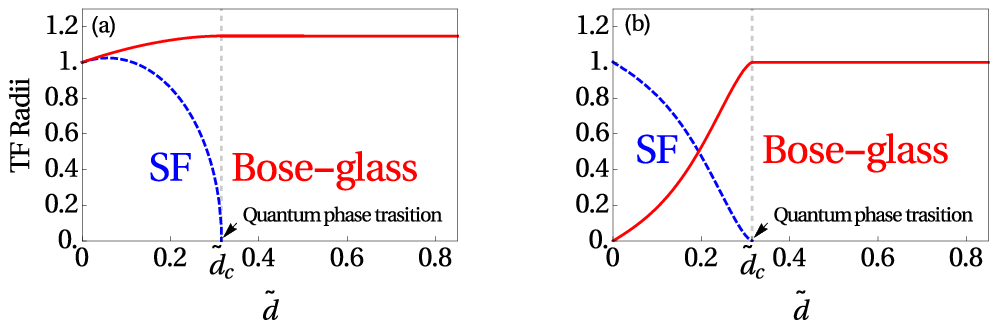}\protect\caption{\label{Radii-TF}(Color online) (a) Condensate radius (dashed, blue)
and cloud radius (solid, red) and (b) fractional number of condensed
particles $N_{0}/N$ (dashed, blue) and in the disconnected local
minicondensates $Q/N$ (solid, red), as a function of disorder strength
$\tilde{d}$. }
\end{figure}

\end{widetext}

Using those parameter values we solve in the superfluid region Eq.~\eqref{17}
for the total density and insert the result into Eqs.~\eqref{15}
and \eqref{16} to get the condensate density and the Bose-glass order
parameter, respectively. This has to be combined with Eq.~\eqref{14}
for the Bose-glass region. After that we fix the chemical potential
$\tilde{\mu}$ using the normalization condition \eqref{18}. The
resulting densities are combined and plotted in Fig.~\ref{TF-Iso}a
for the disorder strength $\ensuremath{\tilde{d}=0.175}$. 

Figure~\ref{TF-Iso}a reveals that, at the condensate radius $\tilde{R}_{{\rm TF1}}$,
a downward jump of the condensate density $\tilde{n}_{0}(\tilde{r})$,
and an upward jump of the Bose-glass order parameter  $\tilde{q}(\tilde{r})$
occur in such a way that the total density $\tilde{n}(\tilde{r})$
remains continuous but reveals a discontinuity of the first derivative.
In the Bose-glass region both the total density and the Bose-glass
parameter coincide and decrease until vanishing at the cloud radius
$\tilde{R}_{{\rm TF2}}$. The TF approximation captures the properties
of the system in both the superfluid and the Bose-glass region but
not in the transition region. This is an artifact of the applied TF
approximation. 

The ratio of the condensate density at the condensate radius $\tilde{n}_{0}(\tilde{R}_{{\rm TF1}})$
with respect to the condensate density at the center of the BEC $\tilde{n}_{0}(0)$
in Fig.~\ref{TF-Iso}b reveals for which range of the disorder strength
the TF approximation is valid. As only a moderate density jump of
about 50\% should be reasonable, our approach is restricted to a dis-order strength of about $\tilde{d}\simeq0.3$. For
a larger disorder strength $\tilde{d}$ one would have to go beyond
the TF approximation and take the influence of the kinetic energy
in Eq.~\eqref{6-2} into account. 

The resulting Thomas-Fermi radii are plotted in Fig.~\ref{Radii-TF}a.
When the disorder strength increases, the condensate radius at first
increases slightly, then decreases until it vanishes, which corresponds
to a quantum phase transition at $\tilde{d}_{{\rm c}}=\frac{2^{1/5}}{\sqrt{3+6\sqrt{3}\cos\frac{\pi}{18}}}\backsimeq0.315$.
This critical value of the disorder strength is obtained by setting
the cloud radius $\tilde{R}_{{\rm TF1}}$ to zero. Thus, superfluidity
is destroyed in our model at a critical disorder strength $\tilde{d}_{{\rm c}}$,
where approximately our TF approximation breaks down. Now we compare
this critical value of the disorder strength with the one obtained
in Refs.~\cite{Nattermann,Natterman2}, where a non-perturbative
approach is used, which investigates energetically shape and size
of the local minicondensates in the disorder landscape. Thus, it is
determined for a decreasing disorder strength once the Bose-glass
phase becomes unstable and goes over in the superfluid phase. In those
references the quantum phase transition for our system parameters
is predicted to occur at the disorder strength value $\tilde{d}=0.115$,
which is of the same order as our $\tilde{d}_{{\rm c}}$.

Contrary to the condensate radius, the cloud radius increases monotonously
with the disorder strength and eventually saturates, so that in the
strong disorder regime the bosonic cloud reaches its maximal radius
of $\underset{\ensuremath{\tilde{d}}\rightarrow\infty}{{\rm lim}}$
$\tilde{R}_{{\rm TF2}}=2^{1/5}\backsimeq1.148$, which is obtained
by inserting the Bose-glass region density \eqref{14} into the normalization
condition \eqref{18}.

The same conclusion can be read off from Fig.~\ref{Radii-TF}b, where
the fractional number of the condensate defined via $N_{0}/N=\frac{15}{2}\int_{0}^{\tilde{R}_{{\rm {TF1}}}}\tilde{r}^{2}\tilde{n}_{0}\left(\tilde{r}\right)d\tilde{r}$
is plotted. We note that $N_{0}/N$ is equal to one in the clean case,
i.e., all particles are in the condensate, then it decreases with
the disorder strength until it vanishes at $\tilde{d}_{{\rm {c}}}$,
marking the end of the superfluid phase and the beginning of the Bose-glass
phase. Conversely, the fraction of atoms in the disconnected minicondensates
$Q/N=\frac{15}{2}\int_{0}^{R_{{\rm {TF2}}}}\tilde{r}^{2}\tilde{q}\left(\tilde{r}\right)d\tilde{r}$
increases with the disorder strength until it reaches its maximum
at $\tilde{d}_{{\rm {c}}}$. Then it remains constant and equals to
one in the Bose-glass phase, since all particles are localized in
the respective minima of the disorder potential.

From Fig.~\ref{TF-Iso}b we conclude that the TF approximation is
valid only in the weak disorder regime, but it is not a good approximation
for intermediate or strong disorder. The TF approximation has a larger
range of validity with respect to the disorder strength in three dimensions
than in one dimension treated in Ref.~\cite{Paper1} due to the fact
that the fluctuations are more violent in lower dimensions. In order
to have a global picture, not only in the presence of weak disorder
but also in the presence of intermediate and strong one, we use in
the following subsection another approximation method to treat the
dirty boson problem: the variational approach.

\subsection{Variational method}

Since the three self-consistency Eqs. \eqref{6-2}, \eqref{5-1},
and \eqref{n-3}, as well as Eq.~\eqref{Q3} are obtained by extremising
the free energy \eqref{F1}, we can apply the variational method in
the spirit of Refs.~\cite{Kleinert1,Kleinert2,Var1,Var2} to obtain
approximate results. In order to be able to compare the variational
results with the analytical ones obtained in the previous subsection,
we use the same rescaling parameters already introduced below Eq.
\eqref{17} for all functions and parameters. To this end, we have
to multiply Eq.~\eqref{F1} with the factor $1/\left(\bar{\mu}\overline{n}R_{{\rm {TF}}}^{3}\right)$
to obtain:
\begin{alignat}{1}
\tilde{\mathcal{F}} & =4\pi\int_{0}^{\infty}d\tilde{r}\tilde{r}^{2}\left\{ -\left[\tilde{q}\left(\tilde{r}\right)+\tilde{n}_{0}\left(\tilde{r}\right)\right]^{2}-\frac{1}{2}\tilde{n}_{0}^{2}\left(\tilde{r}\right)\right.\nonumber \\
 & +\tilde{d}\tilde{Q}_{0}(\tilde{r})\left[\tilde{q}\left(\tilde{r}\right)+\tilde{n}_{0}\left(\tilde{r}\right)\right]-\sqrt{\tilde{n}_{0}\left(\tilde{r}\right)}\Biggl\{\tilde{\tilde{\mu}}+\tilde{d}\tilde{Q}_{0}(\tilde{r})\nonumber \\
 & -2\left[\tilde{q}\left(\tilde{r}\right)+\tilde{n}_{0}\left(\tilde{r}\right)\right]-\tilde{r}^{2}+\left(\frac{\xi}{R_{{\rm {TF}}}}\right)^{2}\frac{1}{\tilde{r}^{2}}\frac{\partial}{\partial\tilde{r}}\left(\tilde{r}^{2}\frac{\partial}{\partial\tilde{r}}\right)\Biggr\}\nonumber \\
 & \times\sqrt{\tilde{n}_{0}\left(\tilde{r}\right)}-2\tilde{d}\left[\tilde{q}\left(\tilde{r}\right)+\tilde{n}_{0}\left(\tilde{r}\right)\right]\nonumber \\
 & \left.\times\sqrt{-\tilde{\mu}'+2\left[\tilde{q}\left(\tilde{r}\right)+\tilde{n}_{0}\left(\tilde{r}\right)\right]+\tilde{r}^{2}-2\tilde{d}\tilde{Q}_{0}(\tilde{r})}\right\} ,\label{F1-3-1}
\end{alignat}
where we have introduced the dimensionless free energy $\tilde{\mathcal{F}}=\mathcal{F}/\left(\bar{\mu}\overline{n}R_{{\rm {TF}}}^{3}\right)$,
the dimensionless chemical potential $\tilde{\mu}'=\mu/\bar{\mu}$,
 and the dimensionless auxiliary function $\tilde{Q}_{0}(\tilde{r})=\frac{1}{\hbar\sqrt{\pi\bar{\mu}}}\left(\frac{2\pi\hbar^{2}}{M}\right)^{3/2}Q_{0}(r)$.

Motivated by the analytical results presented in Fig.~\ref{TF-Iso}a,
we use the following three ansatz expressions for the condensate density
$\tilde{n}_{0}\left(\tilde{r}\right)$, the Bose-glass order parameter
$\tilde{q}\left(\tilde{r}\right)$, and the auxiliary function $\tilde{Q}_{0}(\tilde{r})$:
\begin{equation}
\tilde{n}_{0}\left(\tilde{r}\right)=\alpha e^{-\sigma\tilde{r}^{2}},\label{n0-1}
\end{equation}
\begin{equation}
\tilde{q}\left(\tilde{r}\right)+\tilde{n}_{0}\left(\tilde{r}\right)=\gamma e^{-\theta\tilde{r}^{2}},\label{q-1-1}
\end{equation}
\begin{equation}
\tilde{Q}_{0}(\tilde{r})=2\frac{\tilde{q}\left(\tilde{r}\right)+\tilde{n}_{0}\left(\tilde{r}\right)}{\tilde{d}}-\left(\zeta+\eta\tilde{r}^{2}\right),\label{Q0-2-1}
\end{equation}
where $\alpha$, $\sigma$, $\gamma$, $\theta$, $\zeta$, and $\eta$
denote the respective variational parameters. The parameters $\alpha$
and $\gamma$ are proportional to the number of particles in the condensate
and the total number of particles, while the parameters $\sigma$
and $\theta$ represent the width of the condensate density and the
total density, respectively. Inserting the ansatz \eqref{n0-1}--\eqref{Q0-2-1}
into the free energy \eqref{F1-3-1} and performing the integral yields:

\begin{eqnarray}
\tilde{\mathcal{F}} & = & \pi^{3/2}\left\{ \frac{\sqrt{2}\gamma^{2}}{4\theta^{3/2}}+3\frac{\alpha}{2\sigma^{5/2}}-\frac{\alpha}{8\sigma^{3/2}}\left(8\tilde{\mu}'+\sqrt{2}\alpha\right)\right.\nonumber \\
 &  & \!\!\!\!\!\!\!\!\!\!\!\!\!\!\!\!\!\!+\left(\frac{\xi}{R_{{\rm {TF}}}}\right)^{2}\frac{3\alpha}{2\sqrt{\sigma}}\left.+\tilde{d}\left(\frac{\alpha\zeta}{\sigma^{3/2}}+\frac{3\alpha\eta}{2\sigma^{5/2}}-\frac{\gamma\left(3\eta+2\zeta\theta\right)}{2\theta^{5/2}}\right)\right\} \nonumber \\
 &  & \!\!\!\!\!\!\!\!\!\!\!\!\!\!\!\!\!\!+\frac{2\pi\tilde{d}\gamma\left(\tilde{d}\zeta-\tilde{\mu}'\right)}{\theta\sqrt{1+\tilde{d}\eta}}e^{\frac{\tilde{d}\zeta-\tilde{\mu}'}{2+2\tilde{d}\eta}\theta}K_{1}\left(\frac{\tilde{d}\zeta-\tilde{\mu}'}{2+2\tilde{d}\eta}\theta\right),\label{F1-4-1}
\end{eqnarray}
where $K_{1}\left(s\right)$ represents the modified Bessel function
of second kind.

The free energy \eqref{F1-4-1} has now to be extremised with respect
to the variational parameters $\alpha$, $\sigma$, $\gamma$, $\theta$,
$\zeta$, and $\eta$. Together with the thermodynamic condition $-\frac{\partial\tilde{\mathcal{F}}}{\partial\tilde{\mu}'}=\frac{4}{3}$,
we have seven coupled algebraic equations for seven variables $\alpha$,
$\sigma$, $\gamma$, $\theta$, $\zeta$, $\eta$, and $\tilde{\mu}'$
that we solve numerically.

From all possible solutions we select the physical one with the smallest
free energy, then we insert the resulting variational parameters $\alpha$,
$\sigma$, $\gamma$, and $\theta$ into the variational ansatz \eqref{n0-1}
and \eqref{q-1-1} in order to get the variational total density $\tilde{n}(\tilde{r})$,
the variational condensate density $\tilde{n}_{0}(\tilde{r})$, and
the variational Bose-glass order parameter $\tilde{q}(\tilde{r})$.

\begin{widetext}

\begin{figure}[t]
\includegraphics[width=0.95\textwidth]{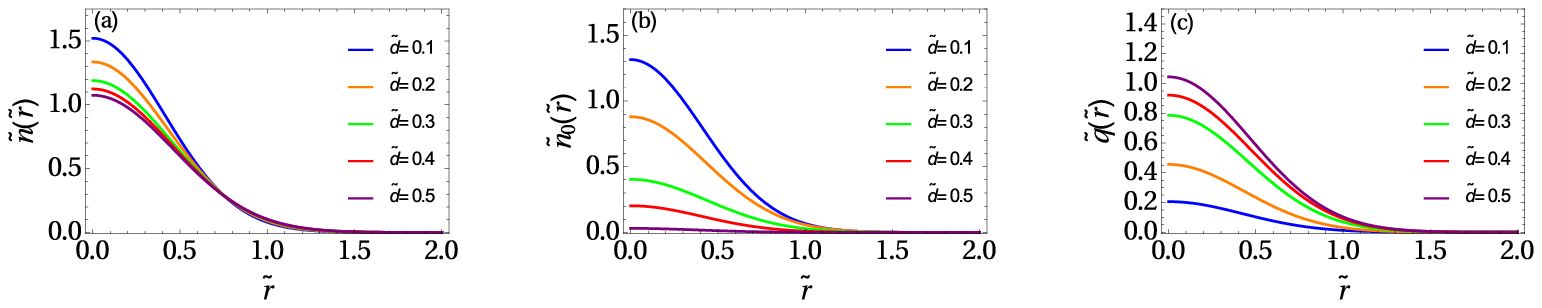}\protect\caption{\label{Densities-var}(Color online) Spatial distribution of: (a)
particle density $\tilde{n}(\tilde{r})$, (b) condensate density $\tilde{n}_{0}(\tilde{r})$,
and (c) Bose-glass order parameter $\tilde{q}(\tilde{r})$ for increasing
disorder strength $\tilde{d}$, from the top to the bottom in the
center in (a) and (b), and from the bottom to the top in (c). }
\bigskip{}
\includegraphics[width=0.8\textwidth]{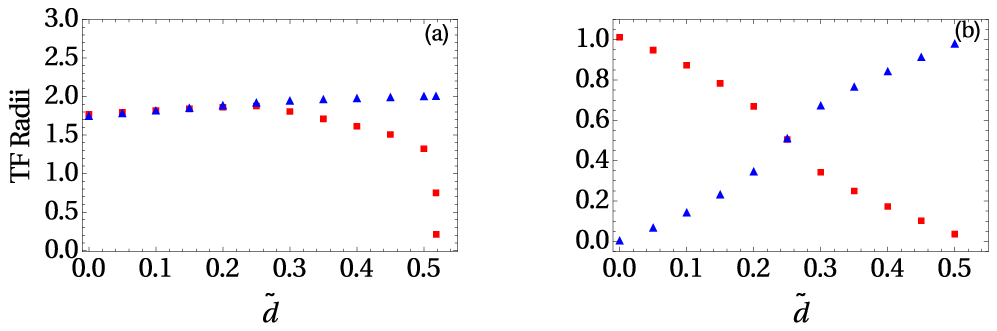}\protect\caption{\label{Radii-Fractions-Var}(Color online) (a) Cloud radius $\tilde{R}_{{\rm {TF2}}}$
(triangle, blue) and condensate radius $\tilde{R}_{{\rm {TF1}}}$
(square, red) and (b) fractional number of condensed particles $N_{0}/N$
(square, red) and fractional number of particles $Q/N$ in the disconnected
local mini-condensates (triangle, blue) as function of disorder strength
$\tilde{d}$.}
\bigskip{}
\includegraphics[width=0.8\textwidth]{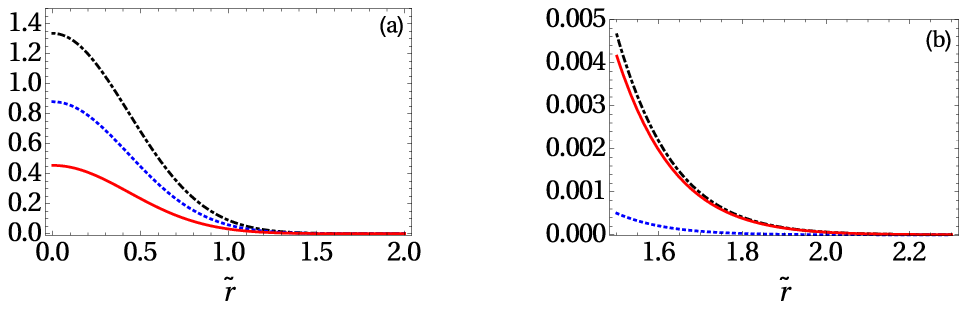}\protect\caption{\label{Bose-glass-Var}(Color online) Spatial distribution of: (a)
particle density $\tilde{n}(\tilde{r})$ (dotted-dashed, black), condensate
density $\tilde{n}_{0}(\tilde{r})$ (dotted, blue), Bose-glass order
parameter $\tilde{q}(\tilde{r})$ (solid, red) and (b) blow-up of
border region for $\tilde{d}=0.35$. }
\end{figure}

\end{widetext}

In Fig.~\ref{Densities-var}a the total density $\tilde{n}(\tilde{r})$
has a Gaussian shape and vanishes at the cloud radius $\tilde{R}_{{\rm {TF2}}}$.
The maximal value of the total density decreases with the disorder
strength. The condensate density $\tilde{n}_{0}(\tilde{r})$ in Fig.~\ref{Densities-var}b
has a similar qualitative behavior as the total density and vanishes
at the condensate radius $\tilde{R}_{{\rm {TF1}}}$. The maximal value
of the condensate density decreases also with the disorder strength.
The response of the condensate density to disorder can be clearly
seen in Fig.~\ref{Radii-Fractions-Var}b, where the fractional number
of condensed particles $N_{0}/N$ is plotted as a function of the
disorder strength. In the clean case all particles are in the condensate,
but, when we increase the disorder strength, more and more particles
leave the condensate until the condensate vanishes at the critical
disorder strength $\tilde{d}_{{\rm {c}}}=0.5183$. 

The Bose-glass order parameter $\tilde{q}(\tilde{r})$ in Fig.~\ref{Densities-var}c
has a similar shape as the two previous densities $\tilde{n}(\tilde{r})$
and $\tilde{n}_{0}(\tilde{r})$. However, when we increase the disorder
strength, the maximal value of the Bose-glass order parameter also
increases. A better understanding of the effect of the dis-order on the local minicondensates can be deduced from
Fig.~\ref{Radii-Fractions-Var}b, where the fractional number of
particles $Q/N$ in the disconnected local minicondensates is zero
in the clean case and then increases with the disorder strength until
reaching the maximal value of one. This means that more and more bosons
go into the local minima of the disorder potential when we increase
the disorder strength. At the critical disorder strength $\tilde{d}_{{\rm {c}}}=0.518$
all particles are in the minicondensates. 

In order to know whether the bosonic cloud contains beside the superfluid
region also a Bose-glass region, we plot the total density $\tilde{n}(\tilde{r})$,
the condensate density $\tilde{n}_{0}(\tilde{r})$, and the Bose-glass
order parameter $\tilde{q}(\tilde{r})$ together in Fig.~\ref{Bose-glass-Var}a
for the disorder strength value $\tilde{d}=0.35$. The blow-up of
the border region in Fig.~\ref{Bose-glass-Var}b shows clearly that
the condensate density vanishes, while the Bose-glass order parameter
still persists, which is the definition of the Bose-glass region.
The cloud radius $\tilde{R}_{{\rm {TF2}}}$ and the condensate radius
$\tilde{R}_{{\rm {TF1}}}$ are conveniently defined by the length,
where the total density and the condensate density are equal to $10^{-4}$,
respectively. Both radii are increasing with the disorder strength
in the weak disorder regime in Fig.~\ref{Radii-Fractions-Var}a.
In the intermediate disorder regime, the cloud radius keeps increasing
monotonously with the disorder strength, while the condensate radius
vanishes at the critical disorder value $\tilde{d}_{{\rm {c}}}=0.518$,
which marks the location of a quantum phase transition. For higher
disorder strengths $\tilde{d}>\tilde{d}_{{\rm {c}}}$ the variational
treatment breaks down as it turns out to have negative solutions for
the condensate density. So with this method it is not possible to
determine if, for stronger disorder, the cloud radius keeps increasing
or remains constant.

\subsection{Comparison between TF approximation and variational results}

\begin{widetext}

\begin{figure}[t]
\includegraphics[width=0.95\textwidth]{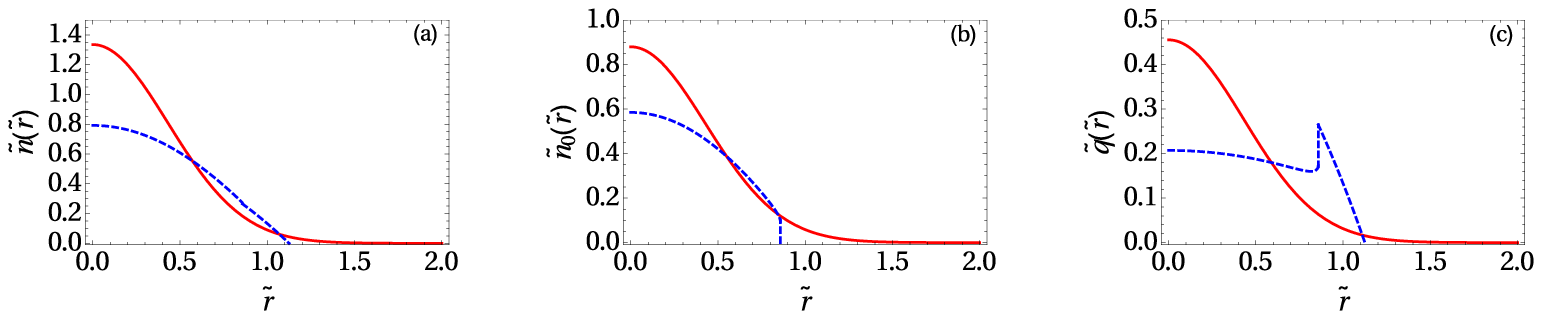}

\protect\caption{\label{all-Iso}(Color online) Spatial distribution of (a) total particle
density $\tilde{n}(\tilde{r})$, (b) condensate density $\tilde{n}_{0}(\tilde{r})$,
and (c) Bose-glass order parameter $\tilde{q}(\tilde{r})$: variational
(solid, red), and analytical (dotted, blue) for $\tilde{d}=0.2$. }
\bigskip{}
\includegraphics[width=0.8\textwidth]{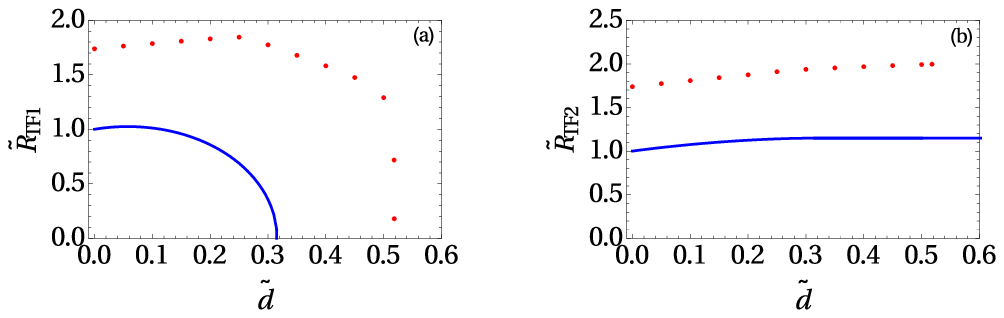}\protect\caption{\label{TFR-Comp-All-Iso}(Color online) Analytical (solid, blue) and
variational (dotted, red) results for (a) condensate radius $\tilde{R}_{{\rm {TF1}}}$
and (b) cloud radius $\tilde{R}_{{\rm {TF2}}}$, as functions of disorder
strength $\tilde{d}$. }
\end{figure}

\end{widetext}

Now we compare the physical quantities obtained via the two different
methods, the TF approximation and the variational approach. We start
with the densities: the total density $\tilde{n}(\tilde{r})$, the
condensate density $\tilde{n}_{0}(\tilde{r})$, and the Bose-glass
order parameter $\tilde{q}(\tilde{r})$ are plotted for the disorder
strength value $\tilde{d}=0.2$ in Fig.~\ref{all-Iso}. We know already
from treating the one-dimensional dirty boson problem in Ref.~\cite{Paper1},
where we also performed extensive numerical simulations, that the
TF approximation describes well the weak disorder regime, while the
variational method is more accurate to describe the intermediate disorder
regime. Based on this conclusion our comparison is here more a qualitative
than a quantitative one. The total densities $\tilde{n}(\tilde{r})$
in Fig.~\ref{all-Iso}a agree qualitatively well. The same can be
said for the condensate density $\tilde{n}_{0}(\tilde{r})$ in Fig.~\ref{all-Iso}b,
except from the jump in the TF-approximated condensate density. For
the Bose-glass order parameter $\tilde{q}(\tilde{r})$ in Fig.~\ref{all-Iso}c
we read off that the TF approximation for the density of the bosons
in the local minima of the disorder potential is maximal at the border
of the trap, but according to the variational result this density
is maximal in the center of the trap. 

The TF-approximated and the variational Thomas-Fermi radii are compared
in Fig.~\ref{TFR-Comp-All-Iso}. In Fig.~\ref{TFR-Comp-All-Iso}a
the variational and the TF-approximated condensate radius $\tilde{R}_{{\rm {TF1}}}$
have the same qualitative behavior, both increase first barely with
the disorder strength $\tilde{d}$ in the weak disorder regime, then
decrease with it in the intermediate disorder regime until they vanish
at the quantum phase transition. Thus, both analytically and variationaly
obtained condensate radii $\tilde{R}_{{\rm {TF1}}}$ indicate the
existence of a quantum phase transition, but at two different values
of the disorder strength, namely $\tilde{d}_{{\rm {c}}}=0.315$ and
$\tilde{d}_{{\rm {c}}}=0.5183$, respectively. The variational quantum
phase transition happens at a larger disorder strength than the TF-approximated
one. Figure~\ref{TFR-Comp-All-Iso}b shows that in the weak disorder
regime, both the variational and the analytical cloud radii $\tilde{R}_{{\rm {TF2}}}$
increase with the disorder strength. In the intermediate disorder
regime the analytical cloud radius remains constant, while the variational
one keeps increasing with the disorder strength. Due to the lack of
information about determine higher disorder strengths $\tilde{d}$,
we can not know if the variational cloud radius keeps increasing even
further or remains constant.

From the discussion above we conclude that the TF approximation and
the variational method are producing similar qualitative results,
contrarily to the one-dimensional case in Ref.~\cite{Paper1}, where
the TF-approximated and the variational results disagree completely.
From studying the one-dimensional dirty boson problem in Ref.~\cite{Paper1},
we can say that the TF approximation produces satisfying results in
the weak disorder regime, while the variational method within the
ansatz \eqref{n0-1}--\eqref{Q0-2-1} is a good approximation to
describe the BEC system in the intermediate disorder regime and has
the advantage of being able to describe the border of the cloud, where
the Bose-glass region is situated and where the TF approximation fails.
Although the variational method does not provide physical results
for larger disorder strengths, its combination together with the TF
approximation for the weak disorder regime covers a significant range
of disorder strengths.

\section{3D Dirty Bosons at Finite Temperature}

In this section we consider the three-dimensional dirty BEC system
to be at finite temperature, so that also the thermal density $n_{{\rm th}}\left(r\right)$
has to be taken into account. After starting with the homogeneous
dirty boson problem, we restrict ourselves first to the trapped clean
case, then we treat the trapped disordered one, both in TF approximation,
where we work out the different densities as well as the respective
Thomas-Fermi radii. This allows us to study the impact of both temperature
and disorder on the distribution of the densities as well as the Thomas-Fermi
radii.

\subsection{Homogeneous case}

\begin{figure}[t]
\includegraphics[width=0.45\textwidth]{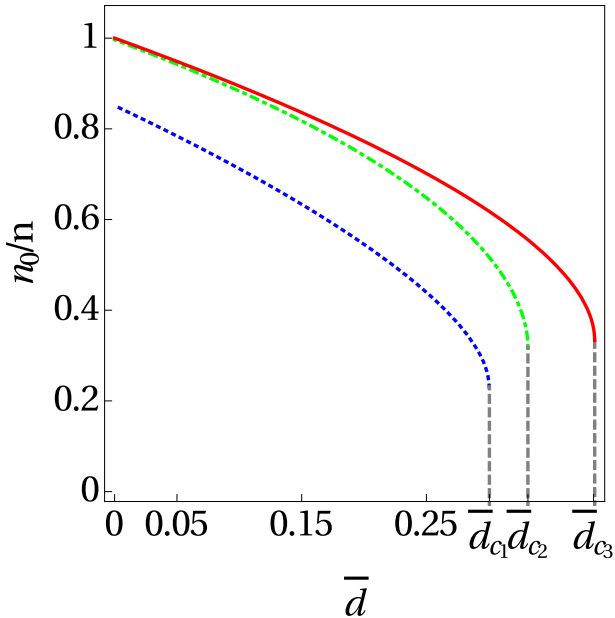} \protect\caption{\label{fig:Homo-3d-T-1}(Color online) Condensate fraction $n_{0}/n$
as function of dimensionless disorder strength $\overline{d}$ for
$\gamma=0.0007$ and $T/T_{{\rm {c}}}^{0}=0.6$ (dotted, blue), $\gamma=0.2366$
and $T/T_{{\rm {c}}}^{0}=0.6$ (dotted-dashed, green), and $T=0$
(solid, red). }
\end{figure}

We start with revisiting the homogeneous case, which was already studied
in Ref.~\cite{Intro-90}, since it is the simplest one. Here we have
$V(r)=0$ and Eqs.~ \eqref{6-2}--\eqref{n-3} reduce in the superfluid
phase to: 
\begin{equation}
gn{}_{0}=\left[\sqrt{-\mu+d^{2}+2gn}+d\right]^{2},\label{6-2-1-1}
\end{equation}

\begin{equation}
q=\frac{dn{}_{0}}{\sqrt{-\mu+d^{2}+2gn}},\label{5-1-2-1}
\end{equation}
\begin{equation}
n_{{\rm th}}=\left(\frac{M}{2\pi\beta\hslash^{2}}\right)^{3/2}\varsigma_{\ 3/2}\left(e^{\beta\ \left(\mu-d^{2}-2gn\right)}\right),\label{3-1-2-1}
\end{equation}

\begin{equation}
n=n_{0}+q+n_{{\rm th}}.\label{n-3-1-1-1}
\end{equation}
Note that we drop in this subsection again the spatial
dependence of all densities due to the homogeneity. From Eqs. \eqref{6-2-1-1}--\eqref{n-3-1-1-1}
we get the following algebraic  equation for the condensate fraction
$n{}_{0}/n$:

\begin{alignat}{1}
 & \left(\frac{n_{0}}{n}\right)^{3/2}-\sqrt{\frac{n_{0}}{n}}+\overline{d}+\left(\frac{T}{\varsigma\left(\frac{3}{2}\right)^{2/3}T_{{\rm {c}}}^{0}}\right)^{3/2}\left(\sqrt{\frac{n_{0}}{n}}-\overline{d}\right)\nonumber \\
 & \times\varsigma_{\ 3/2}\left(e^{-2\frac{T_{{\rm {c}}}^{0}}{T}\varsigma\left(\frac{3}{2}\right)^{2/3}\gamma^{1/3}\ \left[\sqrt{\frac{n_{0}}{n}}-\overline{d}\right]^{2}}\right)=0.\label{Homo-3D-1-1}
\end{alignat}
Here $\overline{d}=\frac{\xi}{\mathcal{L}}$ denotes the dimensionless
disorder strength, $\gamma=na^{3}$ the gas parameter, and $T_{{\rm {c}}}^{0}=\frac{2\pi\hslash^{2}}{Mk_{B}}\left(\frac{n}{\varsigma\left(\frac{3}{2}\right)}\right)^{2/3}$
the critical temperature of the ideal Bose gas, where again $\xi=\frac{\hbar}{\sqrt{2Mgn}}$
stands for the coherence length and $\mathcal{L}=\frac{2\pi\hbar^{4}}{M^{2}D}$
represents the Larkin length \cite{Nattermann,Larkin}. Note that
at zero temperature Eq.~\eqref{Homo-3D-1-1} reduces to Eq.~\eqref{Homo-3D}.
Figure~\ref{fig:Homo-3d-T-1} shows that the condensate fraction
generically decreases with increasing disorder strength $\overline{d}$.
Furthermore, our Hartree-Fock mean-field theory predicts that the
condensate density stops to exist at a critical value $\overline{d}_{{\rm {c}}}$.
We interpret this as a sign that a phase transition occurs in the
homogeneous BEC from the superfluid to the Bose-glass phase. If we
compare in Fig.~\ref{fig:Homo-3d-T-1} the dotted blue line, which
corresponds to a finite temperature, with the solid red line, which
corresponds to the zero-temperature case of Subsection III.A, we observe
that $\overline{d}_{{\rm {c_{1}}}}\simeq0.30<\overline{d}_{{\rm {c_{3}}}}\simeq0.384$
and conclude that the critical disorder strength $\overline{d}_{{\rm {c}}}$
decreases with increasing temperature $T$. Comparing at fixed temperature
the dotted blue line for a weakly interacting $^{87}{\rm {Rb}}$ gas,
which corresponds to the gas parameter of about $\gamma=0.0007$ according
to Ref.~\cite{Weak-Interaction-Stringari}, with the dotted-dashed
green line for a strongly interacting $^{4}{\rm {He}}$, which corresponds
to the gas parameter of about $\gamma=0.2366$ according to Ref.~\cite{Strong-Interaction-Pathria},
yields that $\overline{d}_{{\rm {c_{1}}}}\simeq0.30<\overline{d}_{{\rm {c_{2}}}}\simeq0.331$.
But in order to draw a conclusion how the gas parameter $\gamma$
affects the critical disorder strength, one has to take into account
that it is included in the definition of the dimensionless disorder
strength $\overline{d}=\frac{d}{\sqrt{gn}}$. With this, we conclude
$d_{{\rm {c_{1}}}}>d_{{\rm {c_{2}}}}$, i.e., the critical disorder
strength $d_{{\rm {c}}}$ decreases with increasing the gas parameter
$\gamma$. These findings suggest that a corresponding phase transition
will also appear in the trapped case, which is studied later on in
Subsection IV.G.

\begin{widetext}

\begin{figure}[t]
\includegraphics[width=0.8\textwidth]{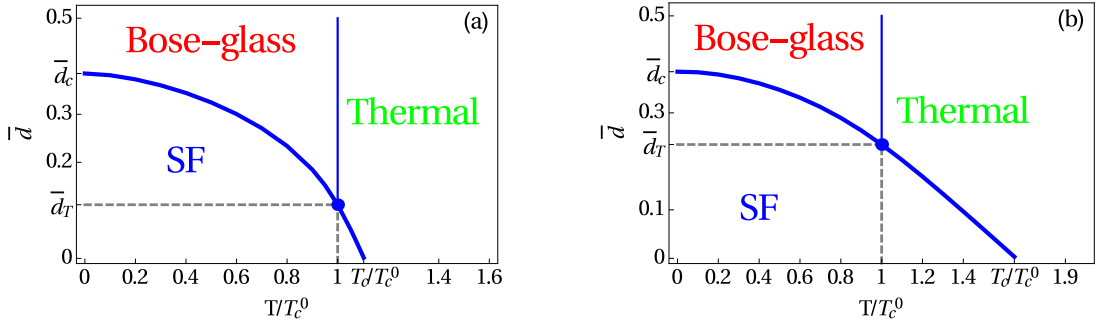} \protect\caption{\label{fig:Phase-Diagram}(Color online) Phase diagram in the disorder
strength-temperature plane for (a) weakly interacting $^{87}{\rm {Rb}}$
gas with $\gamma=0.0007$ and (b) strongly interacting $^{4}{\rm {He}}$
gas with $\gamma=0.2366$. Thick and thin lines represent first order
and continuous phase transitions, respectively. }
\end{figure}

\end{widetext}

To illustrate our results further, we determine where the superfluid,
the Bose-glass, and the thermal phase exist within the phase diagram,
which is spanned by the temperature and the disorder strength. Whereas
this phase diagram was sketched qualitatively in Ref.~\cite{Intro-90},
we determine it here quantitatively in Fig.~\ref{fig:Phase-Diagram}.
This phase diagram follows from solving Eq.~\eqref{Homo-3D-1-1}
together with setting its derivative with respect to $n_{0}/n$ to
zero, i.e.,

\begin{alignat}{1}
 & \frac{3}{2}\sqrt{\frac{n_{0}}{n}}-\frac{1}{2\sqrt{n_{0}/n}}+\frac{1}{2\sqrt{n_{0}/n}}\left(\frac{T}{\varsigma\left(\frac{3}{2}\right)^{2/3}T_{{\rm {c}}}^{0}}\right)^{3/2}\nonumber \\
 & \times\varsigma_{\ 3/2}\left(e^{-2\frac{T_{{\rm {c}}}^{0}}{T}\varsigma\left(\frac{3}{2}\right)^{2/3}\gamma^{1/3}\ \left[\sqrt{\frac{n_{0}}{n}}-\overline{d}\right]^{2}}\right)-\frac{2\gamma^{1/3}}{\sqrt{n_{0}/n}}\nonumber \\
 & \times\sqrt{\frac{T}{\varsigma\left(\frac{3}{2}\right)^{2/3}T_{{\rm {c}}}^{0}}}\left(\sqrt{\frac{n_{0}}{n}}-\overline{d}\right)^{2}\nonumber \\
 & \times\varsigma_{\ 1/2}\left(e^{-2\frac{T_{{\rm {c}}}^{0}}{T}\varsigma\left(\frac{3}{2}\right)^{2/3}\gamma^{1/3}\ \left[\sqrt{\frac{n_{0}}{n}}-\overline{d}\right]^{2}}\right)=0.\label{5.18-1}
\end{alignat}
The phase diagram in Fig.~\ref{fig:Phase-Diagram}a corresponds to
a weakly interacting $^{87}{\rm {Rb}}$ gas, while the phase diagram
in Fig.~\ref{fig:Phase-Diagram}b corresponds to a strongly interacting
$^{4}{\rm {He}}$ gas. The critical disorder strength $\overline{d}_{{\rm {c}}}$
decreases with the temperature $T$. In the clean case $\overline{d}=0$
there is a critical temperature $T_{{\rm {c}}}$ at which the superfluid,
which is stable for $T<T_{{\rm {c}}}$, goes over into the thermal
Bose-gas, which is stable for $T>T_{{\rm {c}}}$. Note that, due to
the weak repulsive interaction, this critical temperature $T_{{\rm {c}}}$
turns out to be larger than the critical temperature of the ideal
Bose gas $T_{{\rm {c}}}^{0}$ by about 
\begin{alignat}{1}
\Delta T_{{\rm {c}}} & =T_{{\rm {c}}}-T_{{\rm {c}}}^{0}\simeq1.3\,\gamma^{1/3}T_{{\rm {c}}}^{0}.\label{5.18}
\end{alignat}

Note that the result \eqref{5.18} is non-trivial as it involves a
resummation of an infrared divergent perturbation series, which has
been worked out on the basis of variational perturbation theory in
Refs.~\cite{5-loops-kleinert,Boris}, and has been confirmed by extensive
MC simulations \cite{Prokof}. For the weakly interacting Bose gas
in Fig.~\ref{fig:Phase-Diagram}a $T_{{\rm {c}}}/T_{{\rm {c}}}^{0}=1.103$,
which agrees well with the result obtained by using formula \eqref{5.18},
where we get $T_{{\rm {c}}}/T_{{\rm {c}}}^{0}\simeq1.115$. The same
can be remarked for the strongly interacting Bose gas in Fig.~\ref{fig:Phase-Diagram}b,
where $T_{{\rm {c}}}/T_{{\rm {c}}}^{0}=1.65$, which agrees well with
the result obtained by using formula \eqref{5.18} $T_{{\rm {c}}}/T_{{\rm {c}}}^{0}\simeq1.796$.
Furthermore, there is a triple point $\overline{d}_{{\rm {T}}}$,
where the three phases coexist and at which $T=T_{{\rm {c}}}^{0}$
and $\mu_{c}=2gn=2g\left(\frac{Mk_{B}T_{{\rm {c}}}^{0}}{2\pi\hslash^{2}}\right)^{3/2}\varsigma\left(\frac{3}{2}\right)$.
So $T_{{\rm {c}}}^{0}$ of the ideal Bose gas turns out to be in our
context the critical temperature for the appearance of the Bose-glass
phase. For $\gamma=0.0007$ we have $\overline{d}_{{\rm {T}}}=0.111$,
while for $\gamma=0.2366$ we obtain $\overline{d}_{{\rm {T}}}=0.234$.
Below the triple-point temperature we have a first-order phase transition
from the superfluid to the Bose-glass phase, while above the triple
point temperature we have a first-order phase transition from the
superfluid to the thermal phase for increasing disorder strength.
Below the triple-point disorder we have for increasing temperature
a first-order phase transition from the superfluid to the thermal
phase, while above the triple point disorder we have a first-order
phase transition from the superfluid to the Bose-glass phase, which
is followed by a second-order phase transition from the Bose-glass
to the thermal phase. At $T=0$ we recover the zero-temperature case,
which was already treated in Subsection III.A.

\subsection{Thomas-Fermi approximation}

After treating the homogeneous case we deal now with the trapped one.
 First we transform Eqs.~\eqref{6-2}--\eqref{n-3} into dimensionless
ones:

\begin{alignat}{1}
\Biggl\{-\tilde{n}_{0}\left(\tilde{r}\right)+\left[\sqrt{-\tilde{\mu}+2\tilde{n}(\tilde{r})+\tilde{r}^{2}}+\tilde{d}\right]^{2}\nonumber \\
-\left(\frac{\xi}{R_{{\rm {TF}}}}\right)^{2}\frac{1}{\tilde{r}^{2}}\frac{\partial}{\partial\tilde{r}}\left(\tilde{r}^{2}\frac{\partial}{\partial\tilde{r}}\right)\Biggr\}\sqrt{\tilde{n}_{0}\left(\tilde{r}\right)} & =0,\label{6-1}
\end{alignat}
\begin{equation}
\tilde{q}\left(\tilde{r}\right)=\frac{\tilde{d}\tilde{n}_{0}\left(\tilde{r}\right)}{\sqrt{-\tilde{\mu}+2\tilde{n}(\tilde{r})+\tilde{r}^{2}}},\label{5-1-1}
\end{equation}
\begin{equation}
\tilde{n}_{{\rm th}}\left(\tilde{r}\right)=\frac{1}{\bar{n}}\left(\frac{M}{2\pi\beta\hslash^{2}}\right)^{3/2}\varsigma_{\ 3/2}\left(e^{\beta\bar{\mu}\ [\tilde{\mu}-2\tilde{n}(\tilde{r})-\tilde{r}^{2}]}\right),\label{3-1-1}
\end{equation}
\begin{equation}
\tilde{n}\left(\tilde{r}\right)=\tilde{n}_{0}\left(\tilde{r}\right)+\tilde{q}\left(\tilde{r}\right)+\tilde{n}_{{\rm th}}\left(\tilde{r}\right).\label{4-1}
\end{equation}

Where dimensionless quantities are as follows: $\tilde{n}_{0}(\tilde{r})=n_{0}(r)/\bar{n}$
denotes the condensate density, $\tilde{q}(\tilde{r})=q(r)/\bar{n}$
the Bose-glass order parameter, $\tilde{n}_{{\rm th}}(\tilde{r})=n_{{\rm th}}(r)/\bar{n}$
the thermal density, $\tilde{n}(\tilde{r})=n(r)/\bar{n}$ the total
density, $\tilde{r}=r/R_{{\rm {TF}}}$ the s radial coordinate, $\tilde{\mu}=(\mu-d^{2})/\bar{\mu}$
the chemical potential, $\tilde{d}=\frac{\xi}{\mathcal{L}}$ the disorder
strength, while $l=\sqrt{\frac{\hbar}{M\varOmega}}$ is the oscillator
length, $R_{{\rm {TF}}}=\sqrt{2\bar{\mu}/M\Omega^{2}}$ the TF cloud
radius at zero temperature, and $\xi=\frac{l^{2}}{R_{{\rm {TF}}}}$
the coherence length in the center of the trap at zero temperature.
The chemical potential in the absence of the disorder at zero temperature
$\bar{\mu}=\frac{15^{2/5}}{2}\Big(\frac{aN}{l}\Big)^{2/5}\hslash\Omega$
is deduced from the normalization condition \eqref{7-1} in the clean
case. We also need to write down the dimensionless equivalent of the
normalization condition $\eqref{7-1}$: 
\begin{equation}
\int_{0}^{\infty}\tilde{n}(\tilde{r})\tilde{r}^{2}d\tilde{r}=\frac{2}{15}.\label{18-1-1}
\end{equation}

For the total density $\tilde{n}(\tilde{r})$, the condensate density
$\tilde{n}_{0}(\tilde{r})$, the Bose-glass parameter $\tilde{q}(\tilde{r})$,
and the thermal density $\tilde{n}_{{\rm th}}(\tilde{r})$ we have
three algebraic equations \eqref{5-1-1}--\eqref{4-1} and one nonlinear
partial differential equation \eqref{6-1}, which is impossible to
solve analytically. Thus we use here again the TF approximation, and
neglect the kinetic term in the self-consistency equation \eqref{6-1},
which becomes in the superfluid region:

\begin{equation}
\tilde{n}_{0}\left(\tilde{r}\right)=\left[\sqrt{-\tilde{\mu}+2\tilde{n}(\tilde{r})+\tilde{r}^{2}}+\tilde{d}\right]^{2},\label{6-1-3}
\end{equation}
where Eqs.~\eqref{5-1-1}--\eqref{4-1} remain the same. Outside
the superfluid region Eq.~\eqref{6-1} is solved by $\tilde{n}_{0}\left(\tilde{r}\right)=0$.

In the following, we treat first the clean case, where we have no
disorder, so as to study only the impact of thermal fluctuations on
the BEC system, and then we treat the general case, where disorder
and temperature occur simultaneously.

\subsection{Clean case}

Even the simpler clean case represents a challenge and has to be treated
in the literature either perturbatively with respect to the interaction
\cite{T-shift1} or fully numerically \cite{Stenholm}. In the clean
case we have no Bose-glass contribution, as we can deduce $\tilde{q}\left(\tilde{r}\right)=0$
from Eq.~\eqref{5-1-1}, but only a thermal contribution $\tilde{n}_{{\rm th}}\left(\tilde{r}\right)$
to the total density $\tilde{n}\left(\tilde{r}\right)$. Therefore,
in this subsection, two different cases have to be distinguished:
in the first one the bosons can be in the condensate or in the excited
states, which corresponds to the superfluid region, while in the second
one all bosons are in the excited states and there is no condensate
any more, so this represents the thermal region. 

Using the Robinson approximation \cite{Robinson,Kleinert2},
\begin{equation}
\!\!\!\varsigma_{\ \nu}\left(e^{x}\right)=\Gamma\left(1-\nu\right)\left(-x\right)^{\nu-1}+\overset{\infty}{\underset{k=0}{\sum}}\frac{x^{k}}{k!}\varsigma\left(\nu-k\right),\, x<0,\!\!\!\label{5.16}
\end{equation}
 for $\nu=3/2$ the TF-approximated Eqs.~\eqref{3-1-1}, \eqref{4-1},
and \eqref{6-1-3} reduce in the superfluid region to:

\begin{alignat}{1}
\tilde{n}_{0}\left(\tilde{r}\right) & \thickapprox\tilde{\mu}-\tilde{r}^{2}-\frac{2g}{\bar{\mu}}\left(\frac{M}{2\pi\beta\hslash^{2}}\right)^{3/2}\Biggl[\Gamma\left(-\frac{1}{2}\right)\sqrt{\beta\bar{\mu}\ \tilde{n}_{0}\left(\tilde{r}\right)}\nonumber \\
 & +\varsigma\left(\frac{3}{2}\right)-\beta\bar{\mu}\ \tilde{n}_{0}\left(\tilde{r}\right)\varsigma\left(\frac{1}{2}\right)\Biggr],\label{234}
\end{alignat}
\begin{equation}
\tilde{n}_{{\rm th}}\left(\tilde{r}\right)=\frac{\tilde{\mu}-\tilde{n}_{0}\left(\tilde{r}\right)-\tilde{r}^{2}}{2},\label{6-1-1-1}
\end{equation}
\begin{equation}
\tilde{n}(\tilde{r})=\tilde{n}_{0}\left(\tilde{r}\right)+\tilde{n}_{{\rm th}}\left(\tilde{r}\right).\label{123-1}
\end{equation}
Equation \eqref{234} represents a quadratic equation with respect
to $\sqrt{\tilde{n}_{0}\left(\tilde{r}\right)}$ and has, thus, two
solutions:

\begin{alignat}{1}
\tilde{n}_{0}\left(\tilde{r}\right) & =\left[-1+2g\beta\left(\frac{M}{2\pi\beta\hslash^{2}}\right)^{3/2}\varsigma\left(\frac{1}{2}\right)\right]^{-2}\nonumber \\
 & \!\!\!\!\!\!\!\!\!\!\!\!\!\!\!\left\{ -\frac{2g}{\bar{\mu}}\left(\frac{M}{2\pi\beta\hslash^{2}}\right)^{3/2}\sqrt{\pi\beta\bar{\mu}}\right.\pm\Biggl\{\frac{4\pi\beta g^{2}}{\bar{\mu}}\left(\frac{M}{2\pi\beta\hslash^{2}}\right)^{3}\nonumber \\
 & \!\!\!\!\!\!\!\!\!\!\!\!\!\!\!-4\left[\frac{-\tilde{\mu}+\tilde{r}^{2}}{2}+\frac{g}{\bar{\mu}}\left(\frac{M}{2\pi\beta\hslash^{2}}\right)^{3/2}\varsigma\left(\frac{3}{2}\right)\right]\nonumber \\
 & \!\!\!\!\!\!\!\!\!\!\!\!\!\!\!\left.\times\left[\frac{1}{2}-\beta g\left(\frac{M}{2\pi\beta\hslash^{2}}\right)^{3/2}\varsigma\left(\frac{1}{2}\right)\right]\Biggr\}^{\frac{1}{2}}\right\} ^{2}.\label{234-1}
\end{alignat}
We choose the one with the positive sign, which corresponds to the
numerical solution without Robinson approximation. We insert this
solution for the condensate density $\tilde{n}_{0}\left(\tilde{r}\right)$
into Eq.~\eqref{6-1-1-1} in order to get the thermal density $\tilde{n}_{{\rm th}}\left(\tilde{r}\right)$,
and the sum of them then represents the particle density $\tilde{n}(\tilde{r})$
according to Eq.~\eqref{123-1}. The condensate radius $\tilde{R}_{{\rm TF1}}$,
which separates the superfluid from the thermal region, is obtained
by setting the derivative of Eq.~\eqref{234} with respect to $\tilde{n}_{0}\left(\tilde{r}\right)$
to zero, i.e., $\frac{\partial\tilde{r}}{\partial\tilde{n}_{0}\left(\tilde{r}\right)}\Biggl|_{\tilde{r}=\tilde{R}_{{\rm TF_{1}}}}=0$.
The resulting condensate density $\tilde{n}_{0}\left(\tilde{R}_{{\rm TF1}}\right)$
is inserted again into equation \eqref{234} in order to get the following
analytical expression for the condensate radius:
\begin{alignat}{1}
\tilde{R}_{{\rm TF1}} & =\Biggl[\tilde{\mu}-\frac{2g}{\bar{\mu}}\left(\frac{M}{2\pi\beta\hslash^{2}}\right)^{3/2}\varsigma\left(\frac{3}{2}\right)\label{RTF1}\\
 & +\frac{1}{\bar{\mu}}\left(\frac{M}{2\pi\beta\hslash^{2}}\right)^{3}\frac{4\pi\beta g^{2}}{1-2g\beta\left(\frac{M}{2\pi\beta\hslash^{2}}\right)^{3/2}\varsigma\left(\frac{1}{2}\right)}\Biggr]^{\frac{1}{2}}.\nonumber 
\end{alignat}

In the thermal region the condensate vanishes, i.e., $\tilde{n}_{0}(\tilde{r})=0$
and $\tilde{n}_{{\rm th}}\left(\tilde{r}\right)=\tilde{n}(\tilde{r})$.
In that case the self-consistency equation \eqref{3-1-1} reduces
to:

\begin{equation}
\tilde{n}\left(\tilde{r}\right)=\frac{g}{\bar{\mu}}\left(\frac{M}{2\pi\beta\hslash^{2}}\right)^{3/2}\varsigma_{\ 3/2}\left(e^{\beta\bar{\mu}\ [\tilde{\mu}-2\tilde{n}\left(\tilde{r}\right)-\tilde{r}^{2}]}\right).\label{4-1-1-1}
\end{equation}
Transcendent Eq.~\eqref{4-1-1-1} contains the polylogarithmic function
$\varsigma_{\ 3/2}$ and, thus, cannot be solved analytically for
$\tilde{n}\left(\tilde{r}\right)$. Furthermore, the Robinson formula
\eqref{5.16} cannot be applied in the thermal region, since it would
yield a diverging density, which is not physical. Thus, the density
of the thermal region \eqref{4-1-1-1} can be treated only numerically.
The cloud radius $\tilde{R}_{{\rm TF2}}$, where the thermal density,
and also as a consequence the total density, vanishes is defined here
conveniently by the length where the thermal density is equal to $10^{-5}$. 

In this subsection we perform our study again for $^{87}{\rm {Rb}}$
atoms with the following experimentally realistic parameters: $N=10^{6}$,
$\Omega=100{\rm \,{Hz}}$, and $a=5.29\,{\rm {nm}}$. For those parameters
the oscillator length is given by $l=2.72{\rm \,{\mu m}}$, the coherence
length in the center of the trap turns out to be $\xi=348.89\,{\rm {nm}}$
and the Thomas-Fermi radius reads $R_{{\rm {TF}}}=21.29{\rm \,{\mu m}}$,
so the assumption $\xi\ll R_{{\rm {TF}}}$ for the TF approximation
is, indeed, fulfilled.

\begin{figure}[t]
\includegraphics[width=0.45\textwidth]{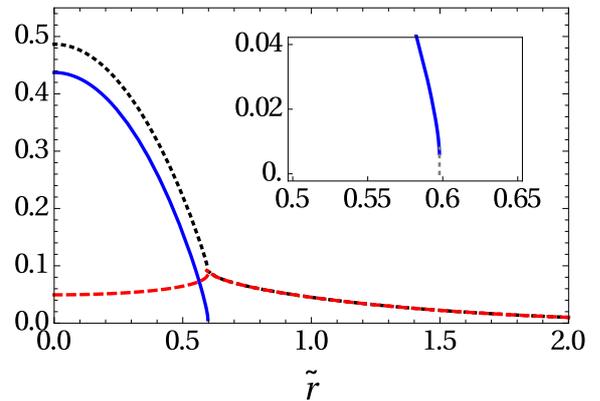}\protect\caption{\label{Densities-T-Clean}(Color online) Total density $\tilde{n}(\tilde{r})$
(dotted, black), condensate density $\ensuremath{\tilde{n}_{0}(\tilde{r})}$
(solid, blue), and thermal density $\tilde{n}_{{\rm th}}\left(\tilde{r}\right)$
(dashed, red) with the blow-up of transition region as a function
of radial coordinate $\ensuremath{\tilde{r}}$ for $T=60~{\rm nK}$
yielding $\ensuremath{\tilde{\mu}=0.566}$.}
\end{figure}

Using those parameter values, we determine the densities of both the
superfluid and thermal region. After that the chemical potential $\tilde{\mu}$
has to be fixed using the normalization condition \eqref{18-1-1},
where the total density $\tilde{n}\left(\tilde{r}\right)$ is the
combination of the total densities from both the superfluid region
and the thermal region. The resulting densities are combined and plotted
in Fig.~\ref{Densities-T-Clean} for the temperature $T=60~{\rm nK}$.

Figure~\ref{Densities-T-Clean} shows that the condensate density
$\ensuremath{\tilde{n}_{0}(\tilde{r})}$ is maximal at the center
of the cloud and decreases when we move away from the center until
the condensate radius $\tilde{R}_{{\rm TF1}}$, where it jumps to
zero. For the chosen parameters the jump is too small to be visible
but it exists as it is shown in the blow-up. The thermal density $\tilde{n}_{{\rm th}}\left(\tilde{r}\right)$
is increasing until reaching its maximum at the condensate radius
$\tilde{R}_{{\rm TF1}}$, and then it decreases exponentially to zero.
The total density $\tilde{n}(\tilde{r})$ is maximal in the trap center
and decreases when one moves away from it, until it vanishes. Note
that in the thermal region the total density $\ensuremath{\tilde{n}(\tilde{r})}$
and the thermal density $\tilde{n}_{{\rm th}}\left(\tilde{r}\right)$
coincide. Although both the condensate density $\ensuremath{\tilde{n}_{0}(\tilde{r})}$
and the thermal density $\tilde{n}_{{\rm th}}\left(\tilde{r}\right)$
are discontinuous at the condensate radius $\tilde{R}_{{\rm TF1}}$,
the total density $\tilde{n}\left(\tilde{r}\right)$ remains continuous
but reveals a discontinuity of the first derivative. We conclude from
Fig.~\ref{Densities-T-Clean} that the condensate is situated in
the trap center, while the bosons in the excited states are located
at the border of the trap.

In order to study how the temperature changes the respective Thomas-Fermi
radii, we plot them in Fig.~\ref{Radii-Fractions-3D-Clean}a as functions
of the temperature $T$. This figure reveals the existence of two
phases: a superfluid phase, where the bosons are either in the condensate
or in the excited\begin{widetext}

\begin{figure}[t]
\includegraphics[width=0.8\textwidth]{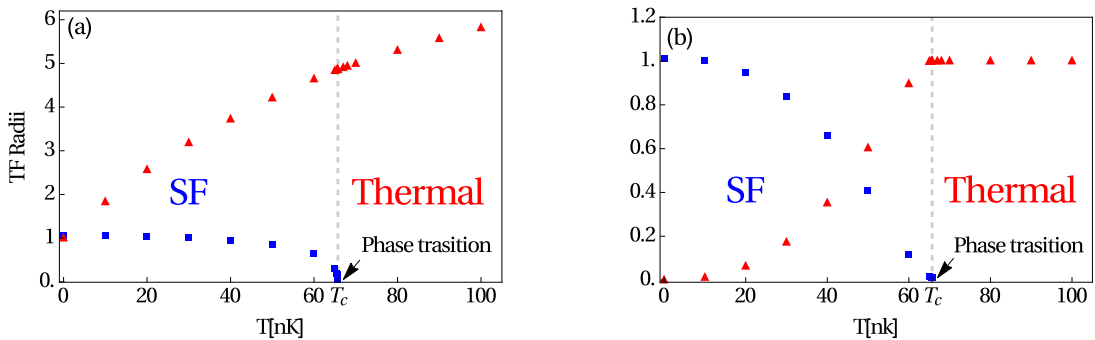}\protect\caption{\label{Radii-Fractions-3D-Clean}(Color online) (a) Condensate radius
$\tilde{R}_{{\rm TF1}}$ (square, blue) and cloud radius $\tilde{R}_{{\rm TF2}}$
(triangle, red) and (b) fractional number of condensed particles $N_{0}/N$
(square, blue) and in the excited states $N_{{\rm {th}}}/N$ (triangle,
red) as a function of temperature $T$. }
\end{figure}
\end{widetext}states, and a thermal phase, where all particles are
in the excited states. The condensate radius $\tilde{R}_{{\rm TF1}}$
decreases with the temperature until it vanishes at the critical temperature
$T_{{\rm {c}}}$ marking the location of the phase transition. The
critical temperature $T_{{\rm {c}}}$ is the solution of the equality
$\tilde{R}_{{\rm TF1}}=0$, i.e., we get from Eq.~\eqref{RTF1} 
\begin{alignat}{1}
\frac{4\pi g^{2}T_{{\rm {c}}}^{2}}{\bar{\mu}k_{B}}\left(\frac{Mk_{B}}{2\pi\hslash^{2}}\right)^{3} & +\left[\tilde{\mu}_{c}-\frac{2g}{\bar{\mu}}\left(\frac{Mk_{B}T_{{\rm {c}}}}{2\pi\hslash^{2}}\right)^{3/2}\varsigma\left(\frac{3}{2}\right)\right]\nonumber \\
 & \!\!\!\!\!\!\!\!\!\!\!\!\!\!\!\!\!\!\!\!\!\!\!\!\!\!\!\!\!\!\!\!\!\!\!\!\!\!\!\!\!\!\!\!\!\times\left[1-\frac{2g\sqrt{T_{{\rm {c}}}}}{k_{B}}\left(\frac{Mk_{B}}{2\pi\hslash^{2}}\right)^{3/2}\varsigma\left(\frac{1}{2}\right)\right]=0,\label{Mu-c}
\end{alignat}
where $\tilde{\mu}_{c}$ is the critical chemical potential at the
phase transition, whose first-order correction follows from Eq.~\eqref{4-1-1-1}
\begin{equation}
\tilde{\mu}_{c}=2\tilde{n}(0)=\frac{2g}{\bar{\mu}}\left(\frac{Mk_{B}T_{{\rm {c}}}}{2\pi\hslash^{2}}\right)^{3/2}\varsigma\left(\frac{3}{2}\right).\label{eq:*-1}
\end{equation}

For the chosen parameters we obtain by solving the system \eqref{Mu-c}
and \eqref{eq:*-1} the values $T_{{\rm {c}}}=65.71~{\rm nK}$ and
$\tilde{\mu}_{c}=0.08$, the former agreeing well with Fig.~\ref{Radii-Fractions-3D-Clean}a.
The critical temperature can be compared with the one given via the
first-order correction \cite{T-shift1,T-shift-2}

\begin{equation}
\frac{T_{{\rm {c}}}-T_{{\rm {c}}}^{0}}{T_{{\rm {c}}}^{0}}=-1.33\frac{a}{l}N^{1/6},\label{Tc}
\end{equation}
where $T_{{\rm {c}}}^{0}=\frac{\hbar\Omega}{k_{B}}\left(\frac{N}{\varsigma\left(3\right)}\right)^{1/3}$
denotes the critical temperature for the non-interacting BEC. Equation
\eqref{Tc} is obtained by inserting Eqs.~\eqref{4-1-1-1} and \eqref{eq:*-1}
into the normalization condition \eqref{18-1-1} and by expanding
the result to first order with respect to the contact interaction
strength $g$. We read off from Eq.~\eqref{Tc} that the repulsive
interaction reduces the critical temperature. For the chosen parameters
the critical temperature of the ideal Bose gas reads $T_{{\rm {c}}}^{0}=71.87~{\rm nK}$.
According to formula \eqref{Tc} the critical temperature for the
interacting case has the value $T_{{\rm {c}}}=70.01~{\rm nK}$, which
is nearly the one obtained above and in Fig.~\ref{Radii-Fractions-3D-Clean}a.
On the other hand, the cloud radius $\tilde{R}_{{\rm TF2}}$ turns
out to increase with the temperature.

In Fig.~\ref{Radii-Fractions-3D-Clean}b the fractional number of
the condensate $N_{0}/N=\frac{15}{2}\int_{0}^{\tilde{R}_{{\rm {TF1}}}}\tilde{r}^{2}\tilde{n}_{0}\left(\tilde{r}\right)d\tilde{r}$
is plotted as a function of the temperature $T$. We note that $N_{0}/N$
is equal to one at zero temperature, i.e., all particles are in the
condensate, then it decreases with the temperature until it vanishes
at $T_{{\rm {c}}}$, marking the end of the superfluid phase and the
beginning of the thermal phase. Conversely, the fractional number
of the particles in the thermal states $N_{{\rm {th}}}/N=\frac{15}{2}\int_{0}^{\tilde{R}_{{\rm {TF2}}}}\tilde{r}^{2}\tilde{n}_{{\rm th}}\left(\tilde{r}\right)d\tilde{r}$,
where $N_{{\rm {th}}}$ is the number of particles in the excited
states, increases with the temperature until being maximal at $T_{{\rm {c}}}$,
and then it remains constant and equals to one in the thermal phase
since all particles are in the excited states.

\begin{figure}[t]
\includegraphics[width=0.45\textwidth]{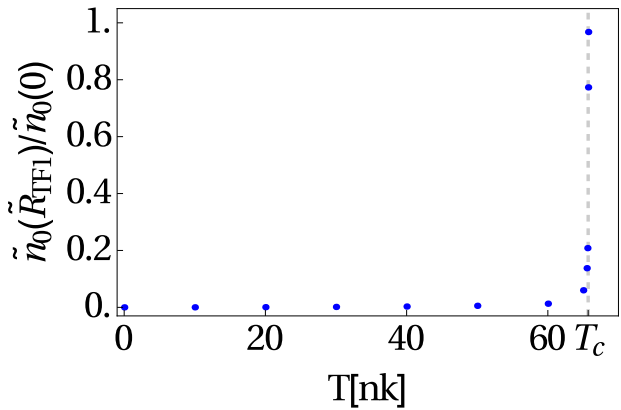}\protect\caption{\label{Mu-Jump-3D-Clean}(Color online) Ratio of $\tilde{n}_{0}(\tilde{R}_{{\rm TF1}})$
and $\tilde{n}_{0}(0)$ as a function of temperature $T$. }
\end{figure}

In order to study for which temperature range the TF approximation
is valid, we plot the ratio of the condensate density at the condensate
radius $\tilde{n}_{0}(\tilde{R}_{{\rm TF1}})$ with respect to the
condensate density at the center of the BEC $\tilde{n}_{0}(0)$ as
a function of the temperature in Fig.~\ref{Mu-Jump-3D-Clean}. We
read off that this ratio is negligible for $T<T_{{\rm {c}}}$ and
has a sudden jump for $T\approx T_{{\rm {c}}}$. This means that the
TF approximation is valid in the superfluid phase but not in the transition
region, where one would have to go beyond the TF approximation and
take the influence of the kinetic energy in Eq.~\eqref{6-1} into
account. 

\subsection{Disordered case}

\begin{figure}[t]
\includegraphics[width=0.45\textwidth]{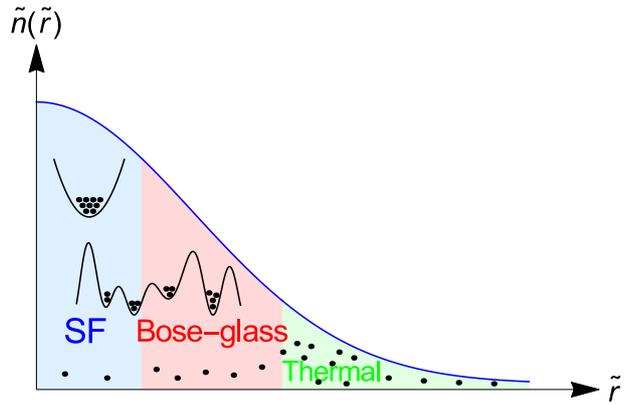}\protect\caption{\label{Logo}(Color online) Illustration for the distribution of bosons
in the superfluid (SF) region, where the condensate density $\tilde{n}_{0}(\tilde{r})$,
the Bose-glass order parameter $\tilde{q}\left(\tilde{r}\right),$
and the thermal density $\tilde{n}_{{\rm th}}\left(\tilde{r}\right)$
contribute to the total density $\tilde{n}(\tilde{r})=\tilde{n}_{0}(\tilde{r})+\tilde{q}\left(\tilde{r}\right)+\tilde{n}_{{\rm th}}\left(\tilde{r}\right)$.
In the Bose-glass region the condensate vanishes and in the thermal
region all particles are in the excited states. }
\end{figure}

In this subsection we consider the BEC system to be in a disordered
landscape as well as at finite temperature. Thus, we investigate now
the effect of both temperature and disorder on the properties of the
system, in particular on the respective densities and Thomas-Fermi
radii. Generically, we have to distinguish three different regions
as illustrated in Fig.~\ref{Logo}: the superfluid region, where
the bosons are distributed in the condensate as well as in the minima
of the disorder potential and in the excited states, the Bose-glass
region, where there are no bosons in the condensate so that all bosons
contribute to the local Bose-Einstein condensates or to the excited
states, and the thermal region, where all bosons are in the excited
states. In the following we analyze the properties of each region
separately. To this end, we have to solve the dimensionless algebraic
Eqs.~\eqref{5-1-1}--\eqref{4-1}, \eqref{6-1-3} and the normalization
condition \eqref{18-1-1}. We start first with the thermal region
and the Bose-glass region, since they are easier to treat, and then
we focus on the superfluid region.

\subsubsection{Thermal region }

In the thermal region only the thermal component contributes to the
total density, so we have $\tilde{n}_{0}(\tilde{r})=\tilde{q}\left(\tilde{r}\right)=0$
and $\tilde{n}_{{\rm th}}\left(\tilde{r}\right)=\tilde{n}(\tilde{r})$.
In this case we just need Eq.~\eqref{3-1-1}, which reduces to

\begin{equation}
\tilde{n}_{{\rm th}}\left(\tilde{r}\right)=\frac{g}{\bar{\mu}}\left(\frac{M}{2\pi\beta\hslash^{2}}\right)^{3/2}\varsigma_{\ 3/2}\left(e^{\beta\bar{\mu}\ [\tilde{\mu}-2\tilde{n}_{{\rm th}}\left(\tilde{r}\right)-\tilde{r}^{2}]}\right),\label{4-1-1-1-1}
\end{equation}
and can only be solved numerically. The cloud radius $\tilde{R}_{{\rm TF3}}$,
which characterizes the end of the thermal region, is determined here
by setting $\tilde{n}_{{\rm th}}\left(\tilde{R}_{{\rm TF3}}\right)=10^{-5}$.

\subsubsection{Bose-Glass region}

In the Bose-glass region the condensate vanishes, i.e., $\tilde{n}_{0}(\tilde{r})=0$,
and we only need the self-consistency equations \eqref{5-1-1}--\eqref{4-1},
which reduce to:

\begin{equation}
\tilde{q}\left(\tilde{r}\right)=\frac{\tilde{\mu}-\tilde{r}^{2}}{2}-\frac{g}{\bar{\mu}}\left(\frac{M}{2\pi\beta\hslash^{2}}\right)^{3/2}\varsigma\left(\frac{3}{2}\right),\label{14-1-1}
\end{equation}
\begin{equation}
\tilde{n}_{{\rm th}}\left(\tilde{r}\right)=\frac{g}{\bar{\mu}}\left(\frac{M}{2\pi\beta\hslash^{2}}\right)^{3/2}\varsigma\left(\frac{3}{2}\right),\label{21}
\end{equation}
\begin{equation}
\tilde{n}(\tilde{r})=\frac{\tilde{\mu}-\tilde{r}^{2}}{2}.\label{22}
\end{equation}
Note that Eq.~\eqref{21} reveals that the thermal density in the
Bose-glass region remains constant, which we consider to be an artifact
of the TF approximation. The Bose-glass radius $\tilde{R}_{{\rm TF2}}$,
which characterizes the end of the Bose-glass region and the beginning
of the thermal region, is determined by setting $\tilde{q}\left(\tilde{R}_{{\rm TF_{2}}}\right)=0$
in Eq.~\eqref{14-1-1}, so we get $\tilde{R}_{{\rm TF2}}=\sqrt{\tilde{\mu}-2\frac{g}{\bar{\mu}}\left(\frac{M}{2\pi\beta\hslash^{2}}\right)^{3/2}\varsigma\left(\frac{3}{2}\right)}$.

\subsubsection{Superfluid region}

In the superfluid region all densities contribute to the total density
and the four algebraic coupled equations \eqref{5-1-1}--\eqref{4-1}
and \eqref{6-1-3} have to be taken into account. We solve them according
to the following strategy. At first we get from Eqs.~\eqref{5-1-1}--\eqref{4-1}
one self-consistency equation for the condensate density $\tilde{n}_{0}(\tilde{r})$:
\begin{alignat}{1}
\left[\sqrt{\tilde{n}_{0}(\tilde{r})}-\tilde{d}\right]^{2}+\tilde{\mu}-\tilde{r}^{2}-\frac{2\tilde{n}_{0}^{3/2}(\tilde{r})}{\sqrt{\tilde{n}_{0}(\tilde{r})}-\tilde{d}}\nonumber \\
-\frac{2g}{\bar{\mu}}\left(\frac{M}{2\pi\beta\hslash^{2}}\right)^{3/2}\varsigma_{\ 3/2}\left(e^{-\beta\bar{\mu}\ \left[\sqrt{\tilde{n}_{0}(\tilde{r})}-\tilde{d}\right]^{2}}\right) & =0.\label{26-1}
\end{alignat}
In the superfluid region we can apply the Robinson formula \eqref{5.16}
for $\nu=3/2$ to approximate Eq.~\eqref{26-1} as 
\begin{alignat}{1}
0= & \left[\sqrt{\tilde{n}_{0}(\tilde{r})}-\tilde{d}\right]^{3}\left[1-2g\beta\left(\frac{M}{2\pi\beta\hslash^{2}}\right)^{3/2}\varsigma\left(\frac{1}{2}\right)\right]\nonumber \\
 & +2\left[g\left(\frac{M}{2\pi\beta\hslash^{2}}\right)^{3/2}\Gamma\left(-\frac{1}{2}\right)\sqrt{\beta/\bar{\mu}}+3\tilde{d}\right]\nonumber \\
 & \times\left[\sqrt{\tilde{n}_{0}(\tilde{r})}-\tilde{d}\right]^{2}+2\tilde{d}^{3}+\left[\sqrt{\tilde{n}_{0}(\tilde{r})}-\tilde{d}\right]\nonumber \\
 & \times\left[2\frac{g}{\bar{\mu}}\left(\frac{M}{2\pi\beta\hslash^{2}}\right)^{3/2}\varsigma\left(\frac{3}{2}\right)+6\tilde{d}^{2}-\tilde{\mu}+\tilde{r}^{2}\right],\label{36}
\end{alignat}
After having solved Eq.~\eqref{36}, we insert the result into the
other algebraic equations. To this end we have to rewrite the other
densities as functions of the condensate density $\tilde{n}_{0}(\tilde{r})$.
From Eqs.~\eqref{5-1-1} and \eqref{6-1-3} we get 
\begin{equation}
\tilde{q}\left(\tilde{r}\right)=\frac{\tilde{d}\tilde{n}_{0}\left(\tilde{r}\right)}{\sqrt{\tilde{n}_{0}(\tilde{r})}-\tilde{d}},\label{37-1}
\end{equation}
and from Eqs.~\eqref{3-1-1} and \eqref{6-1-3} after applying the
Robinson formula \eqref{5.16} for $\nu=3/2$, we obtain:

\begin{alignat}{1}
\tilde{n}_{{\rm th}}\left(\tilde{r}\right) & =\frac{g}{\bar{\mu}}\left(\frac{M}{2\pi\beta\hslash^{2}}\right)^{3/2}\Biggl[\Gamma\left(-\frac{1}{2}\right)\sqrt{\beta\bar{\mu}}\left[\sqrt{\tilde{n}_{0}(\tilde{r})}-\tilde{d}\right]\nonumber \\
 & +\varsigma\left(\frac{3}{2}\right)-\beta\bar{\mu}\ \left[\sqrt{\tilde{n}_{0}(\tilde{r})}-\tilde{d}\right]^{2}\varsigma\left(\frac{1}{2}\right)\Biggr].\label{38}
\end{alignat}

Thus, we have to solve the cubic self-consistency equation for the
condensate density \eqref{36} via the Cardan method and insert the
solution into Eqs.~\eqref{37-1}, \eqref{38}, and \eqref{4-1} in
order to get directly $\tilde{q}\left(\tilde{r}\right)$, $\tilde{n}_{{\rm th}}\left(\tilde{r}\right)$,
and $\tilde{n}(\tilde{r})$, respectively. The cubic Eq.~\eqref{36}
has only one physical solution. To determine the border of the superfluid
region, i.e., the condensate radius $\tilde{R}_{{\rm {TF1}}}$, where
the solution of Eq.~\eqref{36} vanishes and which characterizes
the edge of the superfluid region as well as the beginning of the
Bose-glass region, we determine the first derivative of Eq.~\eqref{36}
with respect to $\tilde{n}_{0}\left(\tilde{r}\right)$, and then we
set $\frac{\partial\tilde{r}}{\partial\tilde{n}_{0}\left(\tilde{r}\right)}\Biggl|_{\tilde{r}=\tilde{R}_{{\rm TF_{1}}}}=0$,
which yields:

\begin{alignat}{1}
3 & \left[\sqrt{\tilde{n}_{0}\left(\tilde{R}_{{\rm TF1}}\right)}-\tilde{d}\right]^{2}\left[1-2g\beta\left(\frac{M}{2\pi\beta\hslash^{2}}\right)^{3/2}\varsigma\left(\frac{1}{2}\right)\right]\nonumber \\
 & +4\left[g\left(\frac{M}{2\pi\beta\hslash^{2}}\right)^{3/2}\Gamma\left(-\frac{1}{2}\right)\sqrt{\beta/\bar{\mu}}+3\tilde{d}\right]\nonumber \\
 & \times\left[\sqrt{\tilde{n}_{0}\left(\tilde{R}_{{\rm TF1}}\right)}-\tilde{d}\right]+\Biggl[2\frac{g}{\bar{\mu}}\left(\frac{M}{2\pi\beta\hslash^{2}}\right)^{3/2}\varsigma\left(\frac{3}{2}\right)\nonumber \\
 & +6\tilde{d}^{2}-\tilde{\mu}+\tilde{R}_{{\rm TF1}}^{2}\Biggr]=0.\label{36-1}
\end{alignat}
 
\begin{figure}[t]
\includegraphics[width=0.45\textwidth]{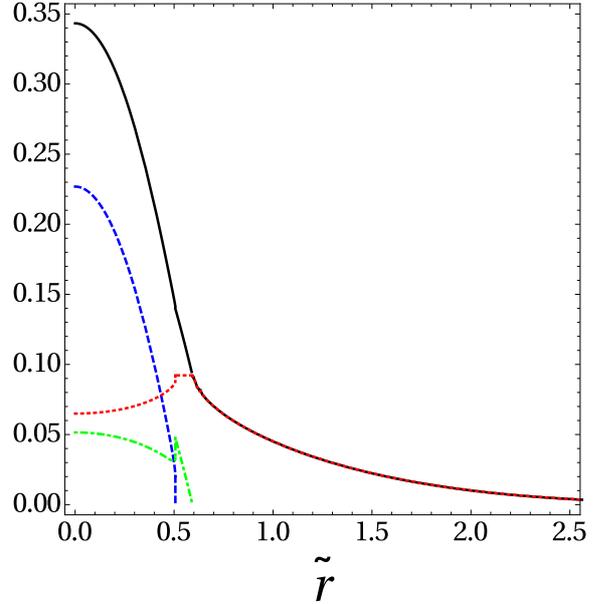}\protect\caption{\label{Densities-3D-T-d}(Color online) Spatial distribution of total
density $\tilde{n}(\tilde{r})$ (solid, black), condensate density
$\ensuremath{\tilde{n}_{0}(\tilde{r})}$ (dashed, blue), Bose-glass
order parameter $\ensuremath{\tilde{q}(\tilde{r})}$ (dotted-dashed,
green), and thermal density $\tilde{n}_{{\rm th}}\left(\tilde{r}\right)$
(dotted, red) as functions of the radial coordinate $\ensuremath{\tilde{r}}$
for $\ensuremath{\tilde{d}=0.088}$. Since $N$ is fixed, $\ensuremath{\tilde{\mu}}$
can be determined and results in $\tilde{\mu}=0.535$. }
\end{figure}
This result we insert back into Eq.~\eqref{36} in order to get the
analytical expression of the condensate radius $\tilde{R}_{{\rm TF1}}$.
As the result is too involved, it is not explicitly displayed here.

\subsection{{\normalsize{}Thomas-Fermi densities}}

Now we perform our study for $^{87}{\rm {Rb}}$ atoms and the same
experimentally realistic parameters as in Subsection IV.C and choose
the temperature to be $T=60~{\rm nK}.$ To this end, we first calculate
the densities in the thermal region, the Bose-glass region, and the
superfluid region. After that we fix the chemical potential $\tilde{\mu}$
using the normalization condition \eqref{18-1-1}, where the total
density $\tilde{n}(\tilde{r})$ is the sum of the densities from all
regions. The resulting densities are plotted in Fig.~\ref{Densities-3D-T-d}.

Figure~\ref{Densities-3D-T-d} shows that the condensate density
$\tilde{n}_{0}(\tilde{r})$ is maximal at the center of the cloud,
then it decreases until reaching its minimum at the condensate radius
$\tilde{R}_{{\rm TF1}}=0.506$. The Bose-glass order parameter $\tilde{q}(\tilde{r})$
is also maximal at the center of the cloud, decreases until the condensate
radius $\tilde{R}_{{\rm TF1}}$ where it jumps upward, then decreases
until reaching its minimum at the Bose-glass radius $\tilde{R}_{{\rm TF2}}=0.588$.
The thermal density $\tilde{n}_{{\rm th}}\left(\tilde{r}\right)$
is behaving differently: it increases until reaching its maximum at
the condensate radius $\tilde{R}_{{\rm TF1}}$, it stays constant
until the Bose-glass radius $\tilde{R}_{{\rm TF2}}$, then it decreases
exponentially to zero. Note that in the thermal region the thermal 

\begin{widetext}

\begin{figure}[t]
\includegraphics[width=0.95\textwidth]{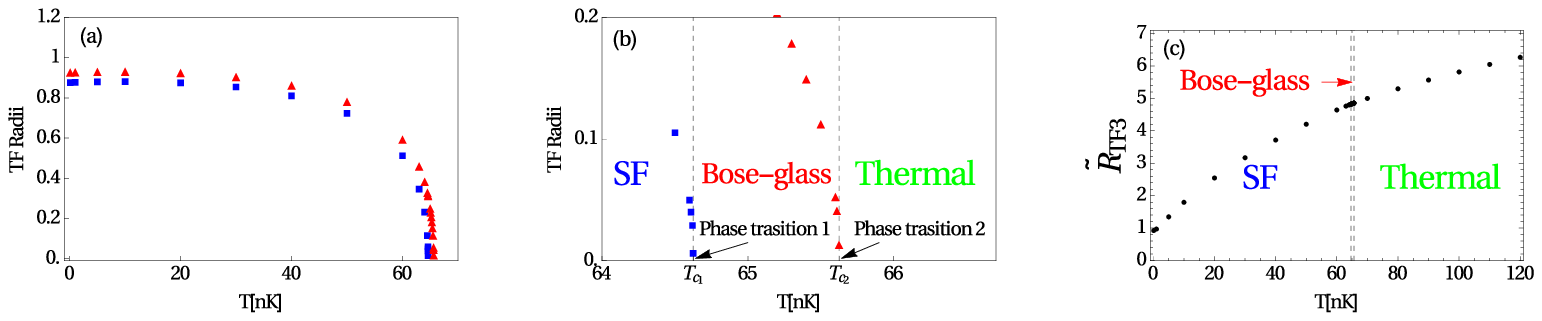}

\protect\caption{\label{Radii-3D-T-Influence}(Color online) (a) Condensate radius
$\tilde{R}_{{\rm TF1}}$ (square, blue) and Bose-glass radius $\tilde{R}_{{\rm TF2}}$
(triangle, red) and (b) blow-up of Bose-glass region (c) cloud radius
$\tilde{R}_{{\rm TF3}}$ (dotted, black) as functions of temperature
$T$.}
\bigskip{}
\includegraphics[width=0.95\textwidth]{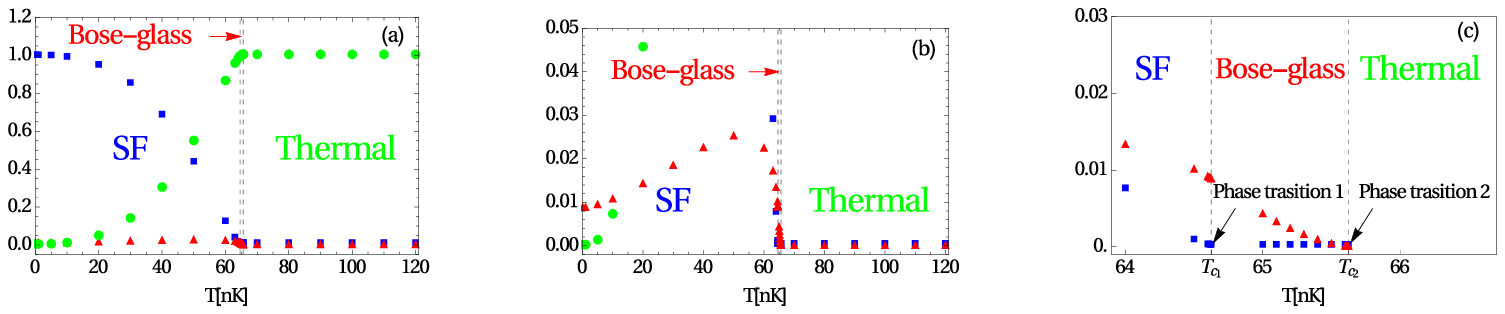}

\noindent \centering{}\protect\caption{\label{Fractions-3D-T-Influence}(Color online) (a) Fractional number
of condensed particles $N_{0}/N$ (square, blue), in disconnected
local minicondensates $Q/N$ (triangle, red), and in excited states
$N_{{\rm {th}}}/N$ (dotted, green) , (b) blow-up of disconnected
local minicondensates $Q/N$, and (c) blow-up of Bose-glass phase
as functions of temperature $\ensuremath{T}$. }
\end{figure}

\end{widetext}density coincides with the total density. The fact
that the thermal density remains constant in the Bose-glass region
is considered to be an artifact of the TF approximation. The total
density $\tilde{n}(\tilde{r})$ is maximal in the center of the trap
and decreases when we move away from the center until it vanishes
at the cloud radius $\tilde{R}_{{\rm TF3}}=4.642$. We note also that,
at the condensate radius $\tilde{R}_{{\rm TF1}}$, a downward jump
of the condensate density $\tilde{n}_{0}(\tilde{r})$, an upward jump
of the Bose-glass order parameter $\tilde{q}(\tilde{r})$, and an
upward jump of the thermal density $\tilde{n}_{{\rm th}}\left(\tilde{r}\right)$
occur in such a way that the total density $\tilde{n}(\tilde{r})$
remains continuous but reveals a discontinuity of the first derivative.
The TF approximation captures the properties of the system within
the superfluid region, the Bose-glass region and the thermal region
but not at the transition point between two regions, namely, between
the superfluid region and the Bose-glass region as well as between
the Bose-glass region and the thermal region. This represents another
artifact of the applied TF approximation. 

In the following we investigate separately the impact of increasing
the temperature $T$ and the disorder strength $\ensuremath{\tilde{d}}$
on the properties of the dirty boson system, namely, the Thomas-Fermi
radii and the fractional number of condensed particles $N_{0}/N$,
in the disconnected local minicondensates $Q/N$, and in the excited
states $N_{{\rm {th}}}/N$.

\subsection{Effects of the temperature }

We start by studying the influence of the temperature on the dirty
boson system. To this end, we fix the disorder strength at $\ensuremath{\tilde{d}}=0.088$
and increase the temperature $T$. The Thomas-Fermi radii are plotted
as functions of the temperature $T$ in Fig.~\ref{Radii-3D-T-Influence}.
Figure ~\ref{Radii-3D-T-Influence}a shows that both the condensate
radius $\tilde{R}_{{\rm {TF1}}}$ and the Bose-glass radius $\tilde{R}_{{\rm {TF2}}}$
decrease with the temperature $T$ until they vanish. The blow-up
in Fig.~\ref{Radii-3D-T-Influence}b reveals that the condensate
radius $\tilde{R}_{{\rm {TF1}}}$ vanishes at $T_{{\rm {\rm {c1}}}}=64.625~{\rm nK}$,
which corresponds to a phase transition from the superfluid to the
Bose-glass. This critical value of the temperature is obtained by
setting the condensate radius $\tilde{R}_{{\rm TF1}}$ to zero. Thus,
superfluidity is destroyed in our model at a critical temperature,
where approximately our TF approximation breaks down. The Bose-glass
radius $\tilde{R}_{{\rm {TF2}}}$ vanishes at $T_{{\rm {\rm {c2}}}}=65.625~{\rm nK}$,
which corresponds to a phase transition from the Bose-glass to the
thermal. This critical value of the temperature is obtained by setting
the Bose-glass radius $\tilde{R}_{{\rm TF2}}$ to zero. The existence
of those two phase transitions means that we are qualitatively above
the triple point introduced for the homogeneous case in Fig.~\ref{fig:Phase-Diagram}.
Note that the difference of both critical temperatures $\triangle T_{{\rm {\rm {c}}}}=T_{{\rm {\rm {c2}}}}-T_{{\rm {\rm {c1}}}}$
is quite small, which is expected, since one can deduce from Eq.~\eqref{4-1-1-1-1}
that the shift $\triangle T$ goes quadratically with the disorder
strength $\ensuremath{\tilde{d}}$, which means that the linear temperature
shift vanishes in agreement with the finding of Ref.~\cite{Timmer}.
Contrary to that, the cloud radius $\tilde{R}_{{\rm TF3}}$ increases
monotonously with the temperature $T$ in Fig~\ref{Radii-3D-T-Influence}c.

\begin{figure}[t]
\includegraphics[width=0.45\textwidth]{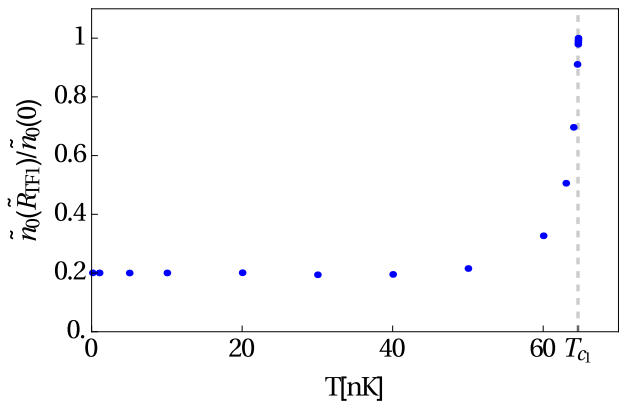}\protect\caption{\label{Jump-3D-T-Influence}(Color online) {\small{}Ratio }$\tilde{n}_{0}(\tilde{R}_{{\rm TF1}})/\tilde{n}_{0}(0)${\small{}
}as a function of temperature $T$. }
\end{figure}

The occupancy fraction of the condensate $N_{0}/N=\frac{15}{2}\int_{0}^{\tilde{R}_{{\rm {TF1}}}}\tilde{r}^{2}\tilde{n}_{0}\left(\tilde{r}\right)d\tilde{r}$,
of the disconnected minicondensates $Q/N=\frac{15}{2}\int_{0}^{\tilde{R}_{{\rm {TF2}}}}\tilde{r}^{2}\tilde{q}\left(\tilde{r}\right)d\tilde{r}$,
and of the excited states $N_{{\rm {th}}}/N=\frac{15}{2}\int_{0}^{\tilde{R}_{{\rm {TF3}}}}\tilde{r}^{2}\tilde{n}_{{\rm th}}\left(\tilde{r}\right)d\tilde{r}$
are plotted in Fig.~\ref{Fractions-3D-T-Influence}a as functions
of the temperature $\ensuremath{T}$. We remark that in the superfluid
phase $N_{0}/N$ decreases with the temperature $\ensuremath{T}$
until vanishing at $T_{{\rm {\rm {c_{1}}}}}$ marking the end of the
superfluid phase and the beginning of the Bose-glass phase as it is
illustrated in the blow-up in Fig~\ref{Fractions-3D-T-Influence}c.
Conversely, in Fig~\ref{Fractions-3D-T-Influence}b we see that $Q/N$
increases with the temperature $T$ until reaching maximum at about
$T=50~{\rm nK}$, then decreases until vanishing at $T_{{\rm {\rm {c2}}}}$
marking the end of the Bose-glass phase and the beginning of the thermal
phase as shown in the blow-up in Fig~\ref{Fractions-3D-T-Influence}c.
In Fig.~\ref{Fractions-3D-T-Influence}a $N_{{\rm {th}}}/N$ increases
starting from zero with the temperature $T$ until being equal to
one at $T_{{\rm {\rm {c_{2}}}}}$, then it remains constant. We conclude
that, by increasing the temperature until $T=50~{\rm nK}$, more and
more particles are leaving the condensate towards the local minicondensates
or the excited states. For the temperature values $50~{\rm nK}<T<T_{{\rm {\rm {c_{1}}}}}$
the particles are leaving both the condensate and the local minicondensates
towards the excited states. When the condensate vanishes at the critical
temperature $T_{{\rm {\rm {c_{1}}}}}$, the particles keep leaving
the local minicondensates towards the excited states until the critical
temperature $T_{{\rm {\rm {c_{2}}}}}$, where all particles are in
the excited states. 

In order to study for which temperature range the TF approximation
is valid, we plot the ratio of the jump of the condensate density
at the Thomas-Fermi condensate radius $\tilde{n}_{0}(\tilde{R}_{{\rm TF_{1}}})$
with respect to the condensate density at the center of the BEC $\tilde{n}_{0}(0)$
as a function of the temperature $T$ in Fig.~\ref{Jump-3D-T-Influence}.
We note that this ratio is negligible for $T<T_{{\rm {c_{1}}}}$ and
has a sudden jump for $T\approx T_{{\rm {c}}}$. This means that the
TF approximation is valid in the superfluid phase but not in the transition
region from the superfluid to the Bose-glass, where one would have
to go beyond the TF approximation and take the effect of the kinetic
energy in Eq.~\eqref{6-1} into account.

\subsection{Disorder effects}

Now we study the influence of the disorder on the dirty boson system.
To this end, we choose the temperature to be $T=60~{\rm nK}$ and
consider an increase of the disorder strength $\ensuremath{\tilde{d}}$. 

\begin{figure}[t]
\includegraphics[width=0.45\textwidth]{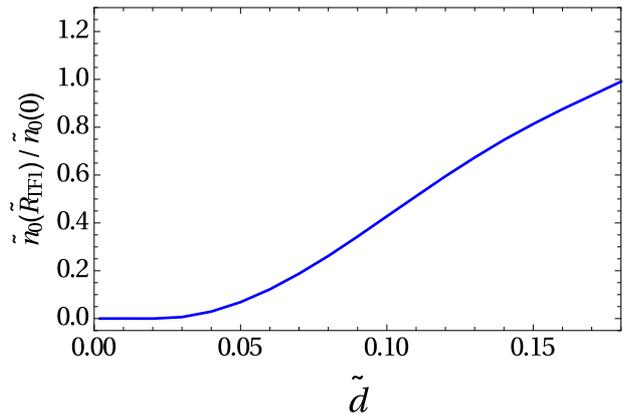}\protect\caption{\label{Mu-Jump-3D-T-d}(Color online) {\small{}Ratio }$\tilde{n}_{0}(\tilde{R}_{{\rm TF1}})/\tilde{n}_{0}(0)${\small{}
}as a function of disorder strength $\ensuremath{\tilde{d}}$. }
\end{figure}

In order to determine for which range of the disorder strength $\ensuremath{\tilde{d}}$
the TF approximation is valid, we plot the ratio of the condensate
density at the Thomas-Fermi condensate radius $\tilde{n}_{0}(\tilde{R}_{{\rm TF_{1}}})$
with respect to the condensate density at the center of the BEC $\tilde{n}_{0}(0)$
as a function of the disorder strength $\ensuremath{\tilde{d}}$ in
Fig.~\ref{Mu-Jump-3D-T-d}. As only a moderate density jump of about
50\% should be reasonable, our approach is restricted to a dimensionless
disorder strength of about $\tilde{d}\simeq0.11$. For a larger disorder
strength $\tilde{d}$ one would have to go beyond the TF approximation
and take the influence of the kinetic energy in \eqref{6-1} into
account. 

\begin{widetext}

\begin{figure}[t]
\includegraphics[width=0.8\textwidth]{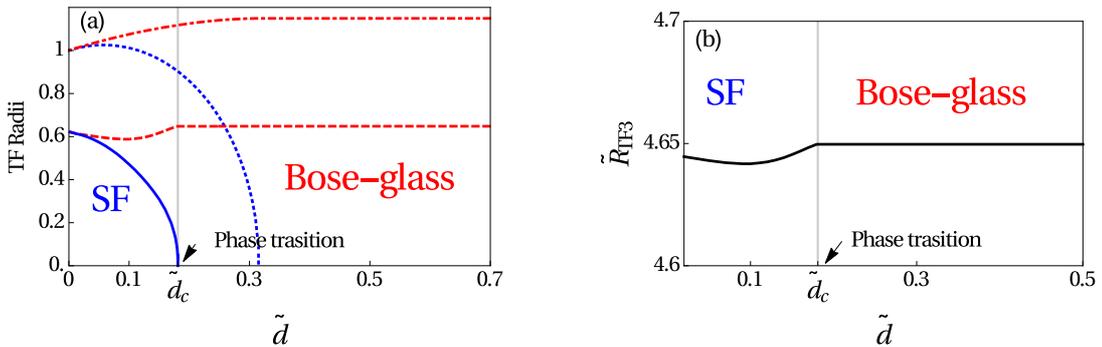}\protect\caption{\label{Radii-3D-T-d}(Color online) (a) Condensate radius $\tilde{R}_{{\rm TF1}}$
at $T=60~{\rm nK}$ (solid, blue) and at $T=0$ (dotted, blue), Bose-glass
radius $\tilde{R}_{{\rm TF2}}$ at $T=60~{\rm nK}$ (dashed, red)
and at $T=0$ (dotted-dashed, red) and (b) cloud radius $\tilde{R}_{{\rm TF3}}$
at $T=60~{\rm nK}$ (solid, black) as functions of disorder strength
$\tilde{d}$.}
\end{figure}

\end{widetext}

The Thomas-Fermi radii are plotted as functions of the disorder strength
$\tilde{d}$ in Fig.~\ref{Radii-3D-T-d}. According to the behavior
of the Thomas-Fermi radii, we distinguish between two different disorder
regimes: the weak disorder regime and the intermediate one. Figure
~\ref{Radii-3D-T-d}a shows that, when the disorder strength $\ensuremath{\tilde{d}}$
increases, the condensate radius $\tilde{R}_{{\rm {TF1}}}$ increases
slightly, then decreases until zero, which corresponds to a phase
transition at about $\tilde{d}_{{\rm {\rm {c}}}}=0.181$. This critical
value of the disorder strength is obtained by setting the cloud radius
$\tilde{R}_{{\rm TF1}}$ to zero. Thus, superfluidity is destroyed
in our model at a critical disorder strength, where approximately
our TF approximation breaks down. Contrarily, the Bose-glass radius
$\tilde{R}_{{\rm {TF2}}}$ decreases when the disorder strength $\tilde{d}$
increases in the weak disorder regime, then increases in the intermediate
disorder regime until the phase transition, then it becomes constant,
so that the bosonic cloud has a maximal Bose-glass radius of $\underset{\ensuremath{\tilde{d}}\rightarrow\infty}{{\rm lim}}$
$\tilde{R}_{{\rm TF2}}=0.647$. Figure~\ref{Radii-3D-T-d}a shows also that in the weak disorder regime the condensate
radius $\tilde{R}_{{\rm {TF1}}}$ and the Bose-glass radius $\tilde{R}_{{\rm {TF2}}}$
coincide, i.e., there is no Bose-glass region, only the superfluid
and the thermal regions exist. Furthermore, comparing the condensate
radius $\tilde{R}_{{\rm {TF1}}}$ and the Bose-glass radius $\tilde{R}_{{\rm {TF2}}}$
at finite temperature with the corresponding ones at zero temperature
reveals that increasing the temperature decreases the critical disorder
strength value $\tilde{d}_{{\rm {\rm {c}}}}$, where the phase transition
is taking place. In Fig~\ref{Radii-3D-T-d}b the cloud radius $\tilde{R}_{{\rm TF3}}$
decreases with the disorder strength $\tilde{d}$ in the weak disorder
regime, then increases with it in the intermediate disorder regime
until becoming constant at the phase transition, so that the bosonic
cloud has a maximal size of $\underset{\ensuremath{\tilde{d}}\rightarrow\infty}{{\rm lim}}$
$\tilde{R}_{{\rm TF3}}=4.649$.
\begin{figure}[t]
\includegraphics[width=0.45\textwidth]{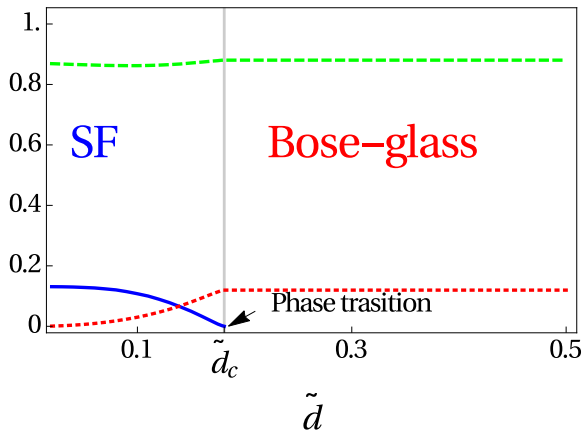}\protect\caption{\label{Fractions-3D-T-d}(Color online) Fractional number of condensed
particles $N_{0}/N$ (solid, blue), in disconnected local minicondensates
$Q/N$ (dotted, red), and in excited states $N_{{\rm {th}}}/N$ (dashed,
green), as functions of disorder strength $\ensuremath{\tilde{d}}$. }
\end{figure}

In Fig.~\ref{Fractions-3D-T-d} the fractional number of the condensate
$N_{0}/N$, in the disconnected minicondensates $Q/N$, and in the
excited states $N_{{\rm {th}}}/N$ are plotted as functions of the
disorder strength $\ensuremath{\tilde{d}}$. We remark that in the
superfluid phase $N_{0}/N$ decreases with the disorder strength $\ensuremath{\tilde{d}}$
until vanishing at $\tilde{d}_{{\rm {\rm {c}}}},$ marking the end
of the superfluid phase and the beginning of the Bose-glass phase.
Conversely, $Q/N$ and $N_{{\rm {th}}}/N$ increase with the disorder
strength $\ensuremath{\tilde{d}}$, i.e., more and more particles
are leaving the condensate towards the local minicondensates or the
excited states. In the Bose-glass phase, both fractions $Q/N$ and
$N_{{\rm {th}}}/N$ remain constant.

\section{Conclusions}

From the presented results we see that for an isotropically trapped
dirty Bose gas the TF approximation provides better description in
three dimensions than in one dimension \cite{Paper1} due to the fact
that the fluctuations are more pronounced in lower dimensions. Additionally,
at zero temperature the respective densities and the Thomas-Fermi
radii obtained via the TF approximation and the variational method
turn out to agree qualitatively well. In particular, a first-order
quantum phase transition from the superfluid phase to the Bose-glass
phase is detected at a critical disorder strength, whose value is
of the same order as the one determined in Refs.~\cite{Nattermann,Natterman2}. 

At finite temperature three regions coexist, namely, the superfluid
region, the Bose-glass region, and the thermal region. Depending on
the parameters of the system, three phase transitions were detected,
namely, from the superfluid to the Bose-glass phase, from the Bose-glass
to the thermal phase, and from the superfluid to the thermal phase.
We have also studied in detail the properties of phase transitions.
The obtained results could be particularly useful for a quantitative
analysis of on going experiments with dirty bosons in three-dimensional
harmonic traps.
\begin{acknowledgments}
The authors gratefully thank Antun Bala\v{z} and Ivana Vasi\'{c}
for discussions. Furthermore, we acknowledge financial support from
the German Academic and Exchange Service (DAAD) and the German Research
Foundation (DFG) via the Collaborative Research Center SFB/TR49 ``Condensed
Matter Systems with Variable Many-Body Interactions''.\end{acknowledgments}

\end{document}